\documentclass[twocolumn]{aastex631}

\usepackage{soul}
\usepackage{comment}

\begin{document}

\title{Star Formation Histories and Stellar Dynamics in the Central Galaxies of 
RX J0820.9+0752, A1835, and PKS 0745-191}

\author{Marie-Jo\"{e}lle Gingras}
\affiliation{Department of Physics and Astronomy, University of Waterloo, Waterloo, ON N2L 3G1, Canada}
\affiliation{Waterloo Centre for Astrophysics, Waterloo, ON N2L 3G1, Canada}

\author{B.R. McNamara}
\affiliation{Department of Physics and Astronomy, University of Waterloo, Waterloo, ON N2L 3G1, Canada}
\affiliation{Waterloo Centre for Astrophysics, Waterloo, ON N2L 3G1, Canada}

\author{Alison L. Coil}
\affiliation{Department of Astronomy and Astrophysics, University of California, San Diego, La Jolla, CA 92093, USA}

\author{Serena Perrotta}
\affiliation{Department of Astronomy and Astrophysics, University of California, San Diego, La Jolla, CA 92093, USA}

\author{Fabrizio Brighenti}
\affiliation{Dipartimento di Fisica e Astronomia, Universitá di Bologna, Via Gobetti 93/2, 40122, Bologna, Italy}
\affiliation{University of California Observatories/Lick Observatory, Department of Astronomy and Astrophysics, Santa Cruz, CA 95064, USA}

\author{S. Peng Oh}
\affiliation{Department of Physics, University of California, Santa Barbara, Santa Barbara, CA 93106, USA}

\author{H.R. Russell}
\affiliation{School of Physics \& Astronomy, University of Nottingham, University Park, Nottingham NG7 2RD, UK}

\author{Wenmeng Ning} 
\affiliation{Department of Astronomy and Astrophysics, University of California, San Diego, La Jolla, CA 92093, USA}
\affiliation{Department of Physics and Astronomy, University of California, Los Angeles, Los Angeles, CA 90095, USA}



\begin{abstract}

We present Keck Cosmic Web Imager observations of stellar populations in three galaxies lying at the centers of cooling flow clusters. All three host rich molecular gas reservoirs and show prominent Balmer absorption from $30-100$ Myr-old stars consistent with long lasting star formation. Two systems, A1835 and PKS 0745$-$191, have spatially extended young stellar populations in their centers with recent star formation rates of 100 M$_{\odot}$ yr$^{-1}$ and 8 M$_{\odot}$ yr$^{-1}$, respectively. In A1835 we uncover a blueshifted clump of young stars moving at high speed with respect to the gas and central galaxy. We suggest these stars formed in a gaseous outflow and have since detached from their natal gas and are now falling inward. This result indicates that star formation is proceeding in a dynamically complex environment shaped by the central galaxy's motion relative to cooling clouds and the feedback from radio jets. In RX J0820.9+0752 intermediate-age stars are found in a filament outside the nucleus with no discernible star formation at the center of the galaxy. All projected galaxies consist of old stellar populations with deep D4000 breaks and lack detectable warm gas. While they may interact gravitationally with the central galaxy, they cannot have donated the upward of $10^{10}~ M_{\odot}$ of molecular gas found in these systems. These results highlight the importance of analyzing spatially resolved stellar kinematics and star formation histories in brightest cluster galaxies, an approach that remains relatively understudied.

\end{abstract}



\section{Introduction} \label{sec:intro}
Clusters of galaxies are home to the most massive elliptical galaxies in the Universe. Referred to as brightest cluster galaxies (BCGs), these galaxies sit at the centers of hot, X-ray-emitting atmospheres, with luminosities ranging between $10^{43}$ and $10^{45}$ erg s$^{-1}$ and temperatures of $10^7–10^8$ K. Under typical central densities of $n_e \gtrsim 10^{-2}$ cm$^{-3}$ and cooling times below one Gyr, hot atmospheres are expected to cool radiatively and condense into molecular clouds, eventually fueling star formation \citep{Fabian1994,Peterson2006,Cavagnolo2008,Pulido2018}.

Multiphase gas and star formation are commonly observed in cool core BCGs. Observations reveal co-spatial hot X-ray gas \citep{Cavaliere1971,Gursky1971,Sarazin1986}, warm ionized nebular gas at $10^4$ K \citep{Heckman1981,Cowie1983,Hu1985,Heckman1989,Crawford1999,Hatch2007,McDonald2010,McDonald2011,Olivares2019,Russell2019}, and cold molecular gas at tens of Kelvin \citep{Lazareff1989,Edge2001,SC2003}. Star formation is common in cooling cores, such as in systems like A1835 and the Phoenix cluster with star formation rates (SFRs) of $\sim100$--$700\ \mathrm{M}_\odot\,\mathrm{yr}^{-1}$ \citep{McDonald2012,McDonald2019,Calzadilla2022}, although cooling cores typically have SFRs of $\sim 1-10$ M$_{\odot}\, \rm{yr}^{-1}$ \citep{Crawford1999,ODea2008,McDonald2011_SF}. These rates are orders of magnitude lower than what is expected from cooling rates inferred from X-ray observations \citep{McNamara1989,Fabian1994,Peterson2006}. This discrepancy—commonly referred to as the ``cooling flow problem"—highlights the need for feedback mechanisms that regulate cooling \citep{Binney1995,McNamara2007,McNamara2012,Fabian2012}.

The most viable heating mechanism capable of injecting enough energy to offset cooling is active galactic nuclei (AGN) feedback, where radio jets launched from the central supermassive black hole (SMBH) interact with the surrounding intracluster medium (ICM). These AGN-driven jets can inflate X-ray cavities by displacing 10$^7 - 10^{10}$ M$_{\odot}$ of hot gas from its surroundings. The cavities then rise buoyantly through the atmosphere, heating it \citep{Churazov2001,Birzan2004,McNamara2007,Rafferty2006,Pope2010,Fabian2012}.

AGN-driven jets, and their associated X-ray cavities, not only heat the galaxy's hot atmosphere but can also trigger localized cooling, eventually leading to star formation \citep{Revaz2008,Brighenti2015,McNamara2016,Donahue2022}. As X-ray cavities rise buoyantly through the galaxy's atmosphere, lower entropy gas from the center of the galaxy is uplifted to higher altitudes.  This displaced gas encounters a new thermodynamic environment and can undergo thermally unstable cooling, resulting in molecular gas condensation and star formation. This process can also result in gas mixing and in metallicity changes as higher metallicity gas is being displaced away from the center \citep{Revaz2008,Simionescu2009,Kirkpatrick2011,Gaspari2013,Li2015,McNamara2016}. 

Multiphase gas and star formation are commonly observed in cool core BCGs in a manner consistent with this model. 
Multiphase gas filaments have been observed in many cooling cores, often trailing rising X-ray cavities \citep{Hu1985,Heckman1989,Crawford1999,Hatch2007,McDonald2010}.

While the nebular and molecular gas phases in BCGs have been studied extensively, stellar populations have received comparatively less attention. Most studies of stars in BCGs have relied on broadband photometry or long-slit spectroscopy to infer stellar ages, metallicities, and SFRs \citep{Loubser2009, Pipino2009, Donahue2015}. 
These methods provide limited spatial and spectral information and cannot resolve the complex interplay between star formation, stellar kinematics, and AGN-driven processes in cluster cores. As a result, it remains unclear how newly formed stars are distributed and whether they retain kinematic signatures of their formation, such as infall, rotation, or lifting.

Our observations address these gaps by providing detailed maps of stellar continuum emission, velocities, and star formation ages across the inner tens of kpc in three BCGs with well-characterized cooling flows and multiphase gas \citep{McNamara2006,McNamara2014,Russell2016,Vantyghem2017,Gingras2024}. We obtained detailed observations of the cores of the BCGs of RX J0820.9+0752, A1835 and PKS 0745-191 using the Keck Cosmic Web Imager \citep[KCWI]{Morrissey2018} which is mounted on the Keck II telescope. We used KCWI's integral field spectroscopy (IFS) abilities to obtain high resolution spatial and spectral observations of the three systems discussed in this paper. The high-quality KCWI data allow us to distinguish the emission from the nebula and the starlight from the host galaxy.

An earlier paper, \cite{Gingras2024}, focused on the nebular gas distribution and kinematics, using the prominent [OII]3726,9 \AA{} emission lines. In this paper we focus on the stellar emission. We analyze the spatial distributions, velocities, and ages of the stars in the central tens of kpc of these galaxies. We also compare the kinematics of the stars to those of the multiphase gas and known AGN feedback features: Atacama Large Millimeter/submillimeter Array (ALMA) observations of the cold molecular gas \citep{McNamara2014,Russell2016,Vantyghem2017,Vantyghem2019}, KCWI observations of the nebular gas \citep{Gingras2024}, and Chandra observations of the X-ray gas and cavities \citep{Hogan2017,Pulido2018,Russell2019,Vantyghem2019}.

In \autoref{sec:Method} we describe our observations and outline the stellar continuum fitting techniques used. In \autoref{sec:Results} we present and analyze flux, velocity and age maps of the stars for each of our targets (RX J0820.9+0752; \autoref{sec:Results:RXJ0820}, A1835; \autoref{sec:Results:Abell 1835}, and PKS 0745-191; \autoref{sec:Results:PKS0745}). In \autoref{sec:Discussion&Conclusion} we summarize the main takeaways from our analysis. Throughout this paper we use a flat $\Lambda$CDM cosmology with $H_0 = 70$ km s$^{-1}$ Mpc$^{-1}$ and $\Omega_{\textrm{m}} = 0.3$.

\section{Observations and data reduction}\label{sec:Method}

The central galaxies discussed in this paper were observed using KCWI. While \cite{Gingras2024} focused on the nebular gas properties, this paper investigates stellar dynamics using the same KCWI observations. The observations and most of the data reduction and analysis are identical for both papers. As the current work focuses on stars, we will refrain from discussing the data reduction and analysis associated with nebular emission, which is discussed in detail in \cite{Gingras2024}.

\subsection{Observations} \label{sec:Method:Observations}

In this paper, we discuss our observations of the stellar emission in the cores of three central galaxies using IFS. \footnote{As part of the same work we observed A262, which contains a circumnuclear disk. Its unusual stellar kinematics are not discussed in this paper.} For all our targets, we used the KCWI blue low-dispersion (BL) grating and medium slicer, resulting in a field of view (FOV) of $16\arcsec.5 \times 20\arcsec.4$ for each pointing, a spaxel scale of $0\arcsec.29 \times 0\arcsec.69$, and a spectral resolution of $\text{R}=1800$. All targets use a detector binning of $2 \times 2$. \autoref{tab:KCWI_obs} lists additional observational parameters for each target. 

As the primary goal of our observations was to study the extended gas nebula in the centers of these galaxies, the number of pointings, their positions, and the total observing area for each target were selected accordingly. Previous optical observations of our targets, both for stellar and nebular emission, were used to tailor our observations to regions of interest.

\begin{deluxetable*}{ccccccccc}
        \tablecaption{KCWI Observations \label{tab:KCWI_obs}}
    \tabletypesize{\scriptsize}
    \tablewidth{0pt} 
    \tablehead{ \colhead{Target} & \colhead{z} &\colhead{Date} & \colhead{Central wavelength}  & \colhead{\# of pointings} & \colhead{Total observing time} & \colhead{Seeing} & \colhead{Total FOV} & \colhead{Total extent} \\ \colhead{} & \colhead{} & \colhead{(DD/MM/YYYY)} &\colhead{(\AA)} &\colhead{} &\colhead{(hrs)} &\colhead{(\arcsec)} &\colhead{($\arcsec \times \arcsec$)} &\colhead{($\textrm{kpc} \times \textrm{kpc}$)}
      }
\colnumbers
\startdata 
RX J0820.9+0752 & 0.1103 & 14/01/2022 & 4650 & 9 & 2.25 & 0.9 & $21 \times 24$ & $44 \times 50$\\
A1835 & 0.2514 & 19/03/2021 & 5300  &  6 & 2 & 0.8 & $35 \times 33$ & $141 \times 134$ \\
PKS 0745-191 & 0.1024 & 12/02/2021 & 4650  & 11 & 3.67 & 1.2 & $26 \times 31$ & $52 \times 61$
\enddata
\tablecomments{Columns: (1) Target name. (2) Redshift (3) Date of KCWI observation. (4) Central wavelength. (5) Number of pointings. (6) Total observing time. (7) Seeing. (8) Total FOV obtained from the different pointings in arcseconds. (9) Same as the previous column but in kpc. This table is a subset of Table 2 presented in \cite{Gingras2024}.
}
\end{deluxetable*}

\subsection{Data Reduction}\label{sec:Methods:Data_reduction}

The data were reduced using the KCWI data extraction and reduction pipeline (KDERP) in IDL \citep{KDERP}. KDERP includes bias subtraction, gain corrections, cosmic ray removal, atmospheric refraction correction, scattered light and sky subtraction, and flux calibration using a standard star. The scattered light and sky subtraction steps of the data reduction require blank sky images, which are obtained by manually masking emission sources from the images. For each pointing, the corresponding blank sky image was subtracted from the raw observations.
Finally, for each target, the various pointings were resampled onto a grid with spaxel sizes of $0\arcsec.29 \times 0\arcsec.29$ and summed together to form a mosaic, using routines from the IFSRED library \citep{Rupke2014a}.

\subsection{Stellar Continuum Fitting}\label{sec:Methods:Stellar_fit}

Each spaxel in the reduced mosaics was fitted using the IFSFIT IDL library from \cite{Rupke2014b}. All spectra are fitted using both the Penalized Pixel-Fitting (PPXF) method \citep{PPXF} to fit the stellar continuum, and MPFIT \citep{MPFIT} to fit the emission lines.

In this work, we focus on the stellar continuum fits, where emission lines and known sky lines are masked. We fitted the resulting stellar continuum using the PPXF method from \cite{PPXF}. This method uses a linear combination of stellar population synthesis (SPS) models at solar metallicity \citep{Choi2016,Byrne2023} and Legendre polynomials. The stellar population models used throughout this paper consist of a set of 43 SPS templates, all with stellar metallicity, which vary in stellar ages between 1 Myr and 9 Gyr. Legendre polynomials are used when fitting the continuum to account for issues remaining after the data reduction process such as atmospheric effects, scattered light, and/or residual features in the stellar continuum. The contribution of each SPS template is taken to be its fractional contribution to the total stellar continuum at 5000 \AA{} (rest-frame).

Before fitting each spaxel in our reduced mosaics, we had to define the systemic velocity of each system, which is used as the 0 km s$^{-1}$ speed throughout our analysis. To create a high-quality spectrum from which we measured the systemic velocity, the spaxels with the brightest stellar emission were summed together. The resulting stellar continuum was fitted using the PPXF method previously described. In this case, we used Legendre polynomials of order 30. The stellar velocity obtained from this fit is taken to be the systemic velocity.

However, the stellar continuum fit requires an initial guess for the systemic velocity, which we took from \cite{Russell2019}. Since the measured stellar velocity depends on the accuracy of the initial guess, we iteratively measured the stellar velocity using the previous result as our initial guess. This process stopped when the uncertainty on the fitted stellar velocity exceeded the change between the values from successive iterations by at least an order of magnitude. 

Once a systemic velocity is defined, we fitted each spaxel in our reduced mosaic using the PPXF \citep{PPXF} method previously discussed with a Legendre polynomial of order 15.

Both our emission line fits and our continuum fits are corrected for extinction. For the nebular gas, the color excess $E(B-V)_{\text{gas}}$ is measured from the Balmer decrement, using the ratio of the total fluxes (per spaxel) of H$\beta$ to H$\gamma$. We used the extinction curve from \cite{Cardelli1989} for Case B recombination at $10^4$ K and $R_V=3.1$. When either the H$\beta$ or H$\gamma$ flux could not be observed, the median $E(B-V)_{\text{gas}}$ value was used instead. The stellar broadband color excess for the stars, $E(B-V)_*$, was obtained from the PPXF fit, which uses the \cite{Calzetti2000} attenuation curve with $R_V=4.05$. We obtained the value for $E(B-V)_*$ and estimated the 1$\sigma$ uncertainties using Monte Carlo simulations.

Stellar masses and star formation rates were estimated from the stellar continuum fits. To calculate stellar masses, we used B-band mass-to-light ratios ($M/L_B$) from the \cite{B&C2003} stellar population models, which assume solar metallicity and a Salpeter initial mass function. Observed spectra were convolved with a B-band transmission curve using the synphot Python package \citep{synphot} to derive the B-band luminosity. From the continuum fits, we know the relative contribution of each SPS template at 5000 \AA. Since B-band luminosities and mass-to-light ratios were used in these calculations, we took the ratio of the flux at 4420 \AA{} and 5000 \AA{} to get the relative contribution of each template at 4420 \AA, which is the central wavelength of the Johnson B-band. The B-band luminosity contribution of each SPS template was then combined with the corresponding $M/L_B$ to calculate the stellar mass associated with each template. The total stellar mass and SFRs were computed by summing the mass contributions over all templates, using the age of each population to estimate the average SFR over the relevant time interval.

\section{Results}\label{sec:Results}

Using KCWI integral field spectroscopy we are able to map both the nebular and stellar emission in the cores of our targets. We focus here on the stellar properties, while \cite{Gingras2024} focused on the kinematics of the nebular gas. In this section we analyze the stellar spatial distributions, velocities, and ages in the cores of the central galaxies of RX J0820.9+0752 (\autoref{sec:Results:RXJ0820}), A1835 (\autoref{sec:Results:Abell 1835}), and PKS 0745-191 (\autoref{sec:Results:PKS0745}).

\subsection{RX J0820.9+0752}\label{sec:Results:RXJ0820}
\begin{figure}
    \centering
    \includegraphics[width=\columnwidth]{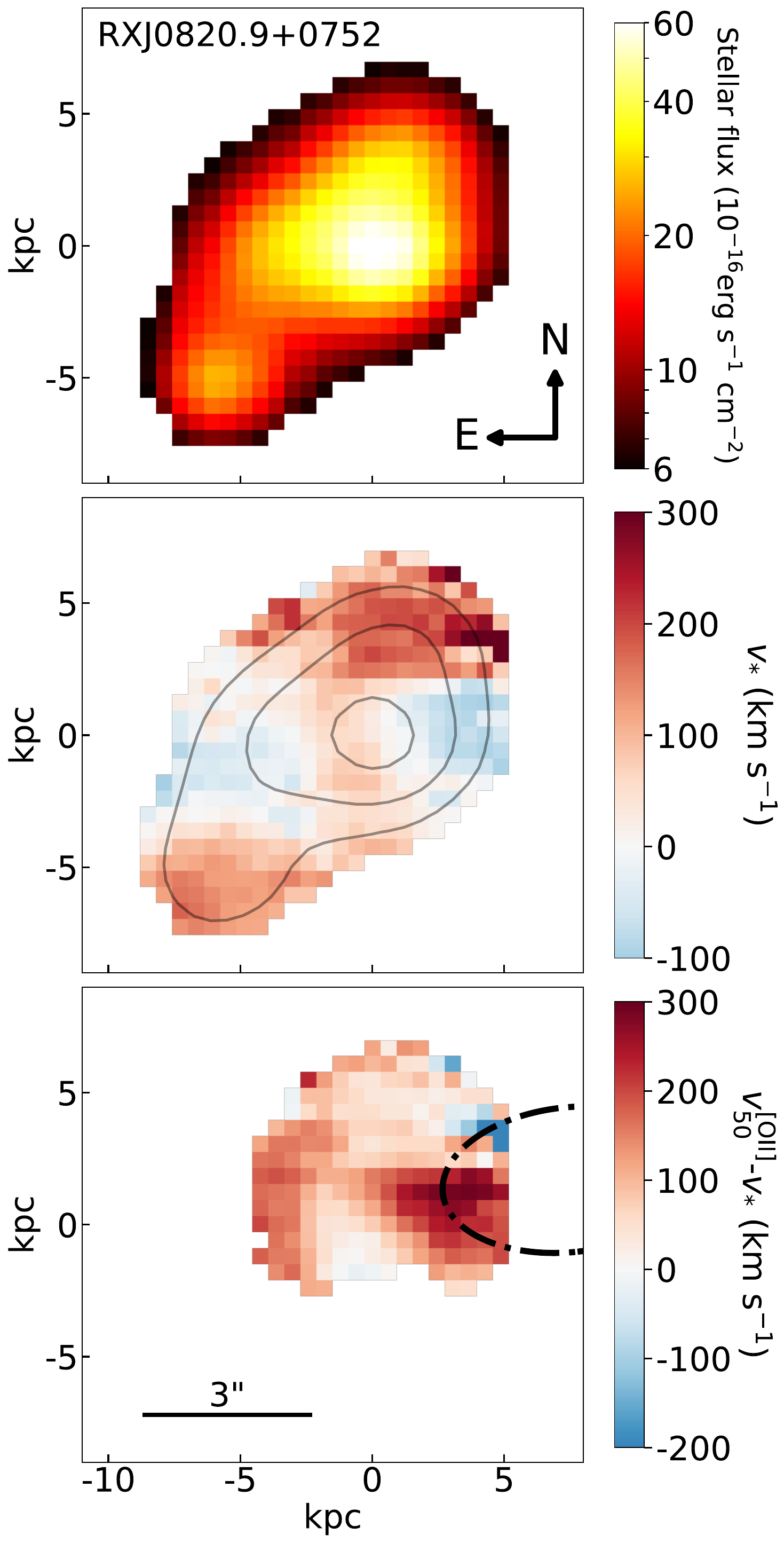}
    \caption{Maps of stellar properties of the central galaxy in  RX J0820.9+0752. Top: Integrated stellar flux map for rest-frame wavelengths of $3300-5100$ \AA. Only spaxels with stellar fluxes of at least $5 \times 10^{-16}$ erg s$^{-1}$ cm$^{-2}$ are shown. Center: Median stellar velocity map. The grey lines show stellar flux contours corresponding to 20\%, 40\% and 80\% of the maximum stellar flux per spaxel. Bottom: Velocity of the nebular gas with respect to the stellar velocity. Unless specified otherwise, the velocity of the nebular gas is taken to be the total [OII]3726,9 \AA{} emission line median velocity ($v_{50}^{\textrm{[OII]}}$). The black dash-dotted ellipse shows the position of the X-ray cavity observed by \cite{Vantyghem2019}.}
    \label{fig:RXJ0820:flux;vstel;v50-vstel}
\end{figure}

\begin{figure*}
    \centering
    \includegraphics[width=\textwidth]{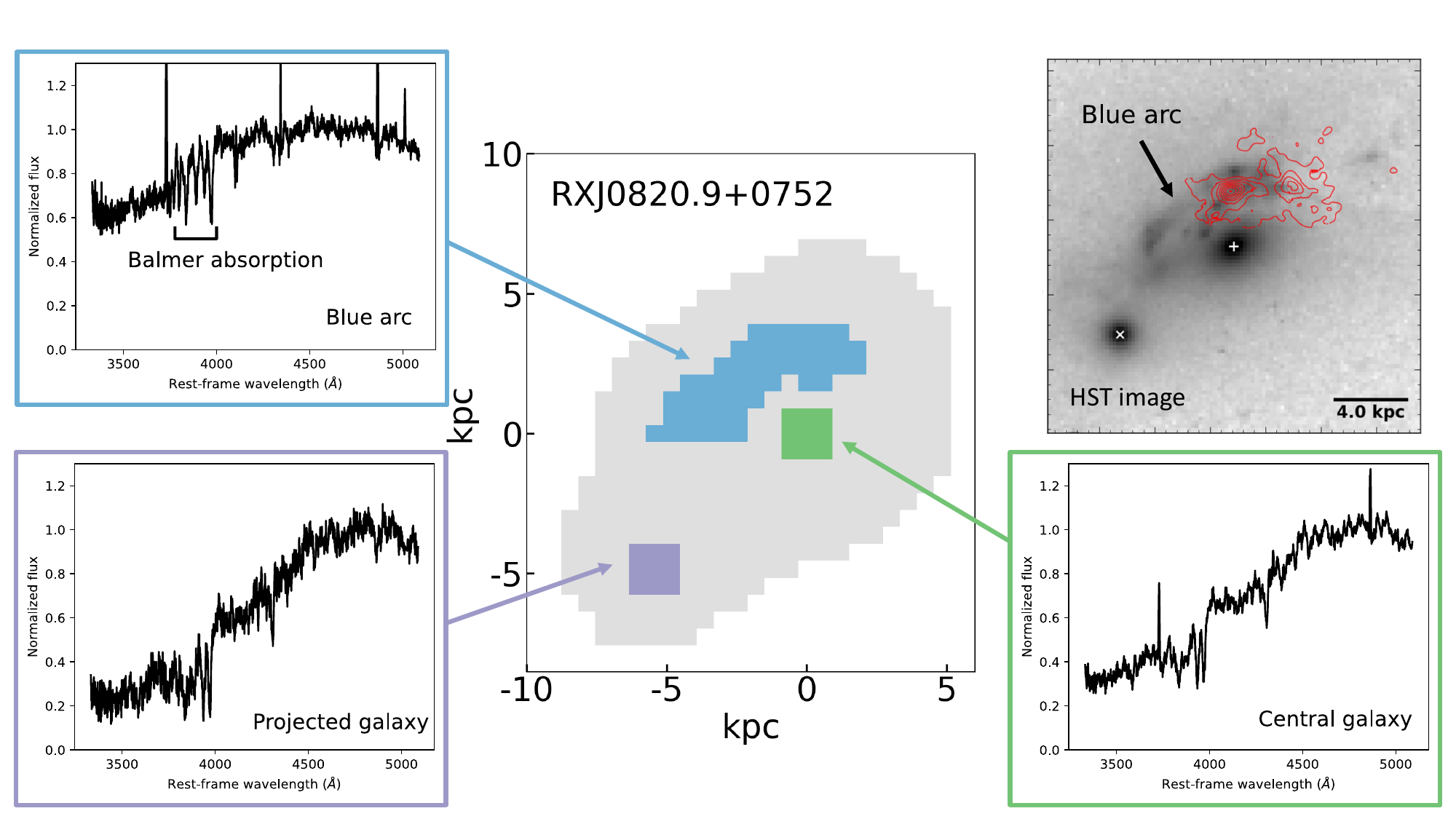}
    \caption{Spectra of three regions in  RX J0820.9+0752: 0.5" (radius) around the nucleus (green), 0.5" around the nucleus of the secondary galaxy (purple) and the arc of young stars previously seen in Hubble observations \citep{Bayer-Kim2002,Vantyghem2017}. The upper right panel shows the HST image of RX J0820.9+0752 as shown in \cite{Vantyghem2017}. The red contours show the distribution of CO(3-2) from ALMA.
    The grey region in the central panel shows the footprint of the stellar continuum as plotted in the top panel of \autoref{fig:RXJ0820:flux;vstel;v50-vstel}. The colored regions in the 2D map show the spaxels included in each integrated spectrum. The vertical axis of the spectra shows the normalized flux where the flux is normalized by the average flux between 4700 ${\rm\AA}$ and 4800 $\rm \AA$, allowing us to compare the continuum shapes of different regions.}
    \label{fig:regions:RXJ0820}
\end{figure*}

\begin{figure}
    \centering
    \includegraphics[width=\columnwidth]{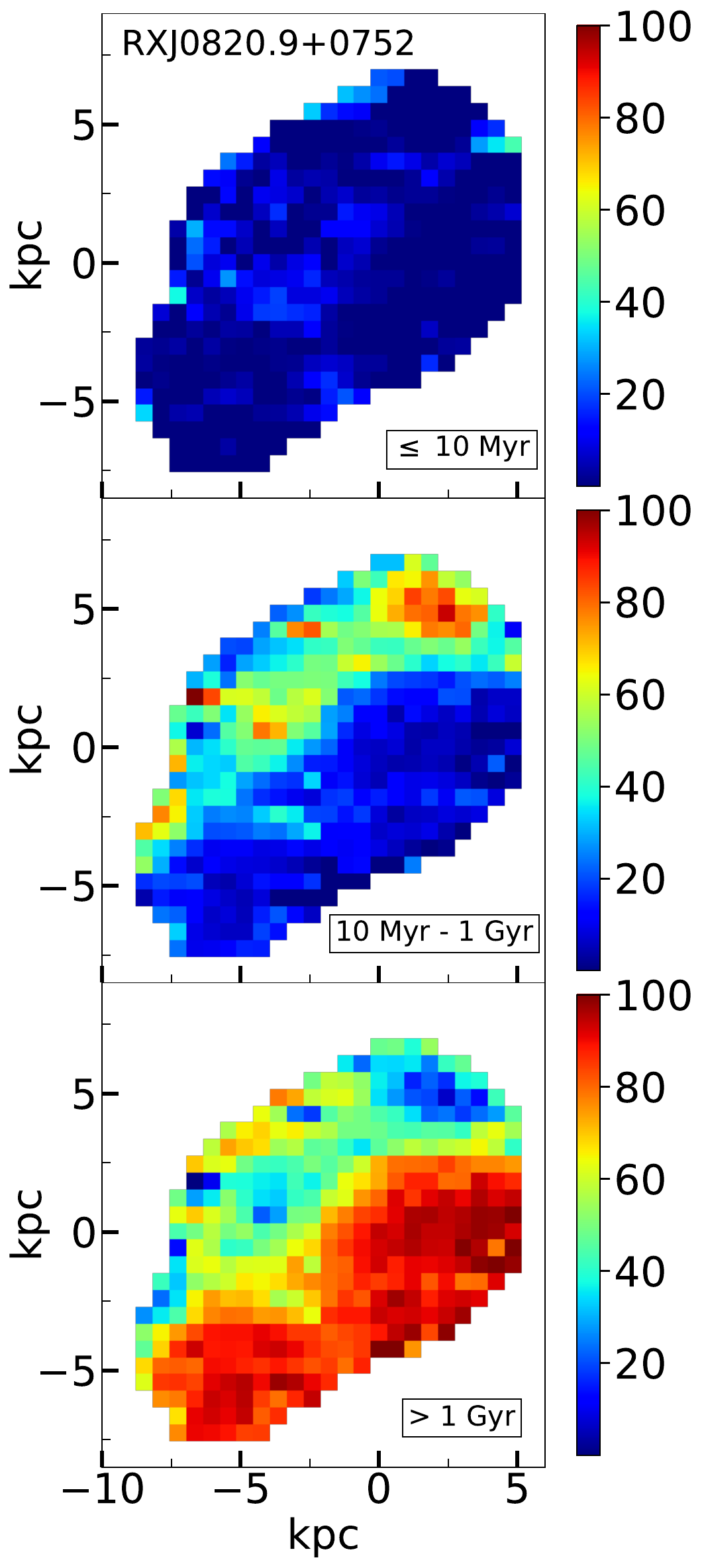}
    \caption{Distributions of stellar ages in  RX J0820.9+0752. The colorbars show the fraction of the stellar flux emitted by young ($\leq$ 10 Myr), intermediate-age (10 Myr $-$ 1 Gyr) and old ($>$ 1 Gyr) stellar populations.}
    \label{fig:RXJ0820:Ages_maps}
\end{figure}

\begin{figure}
    \centering
    \includegraphics[width=\columnwidth]{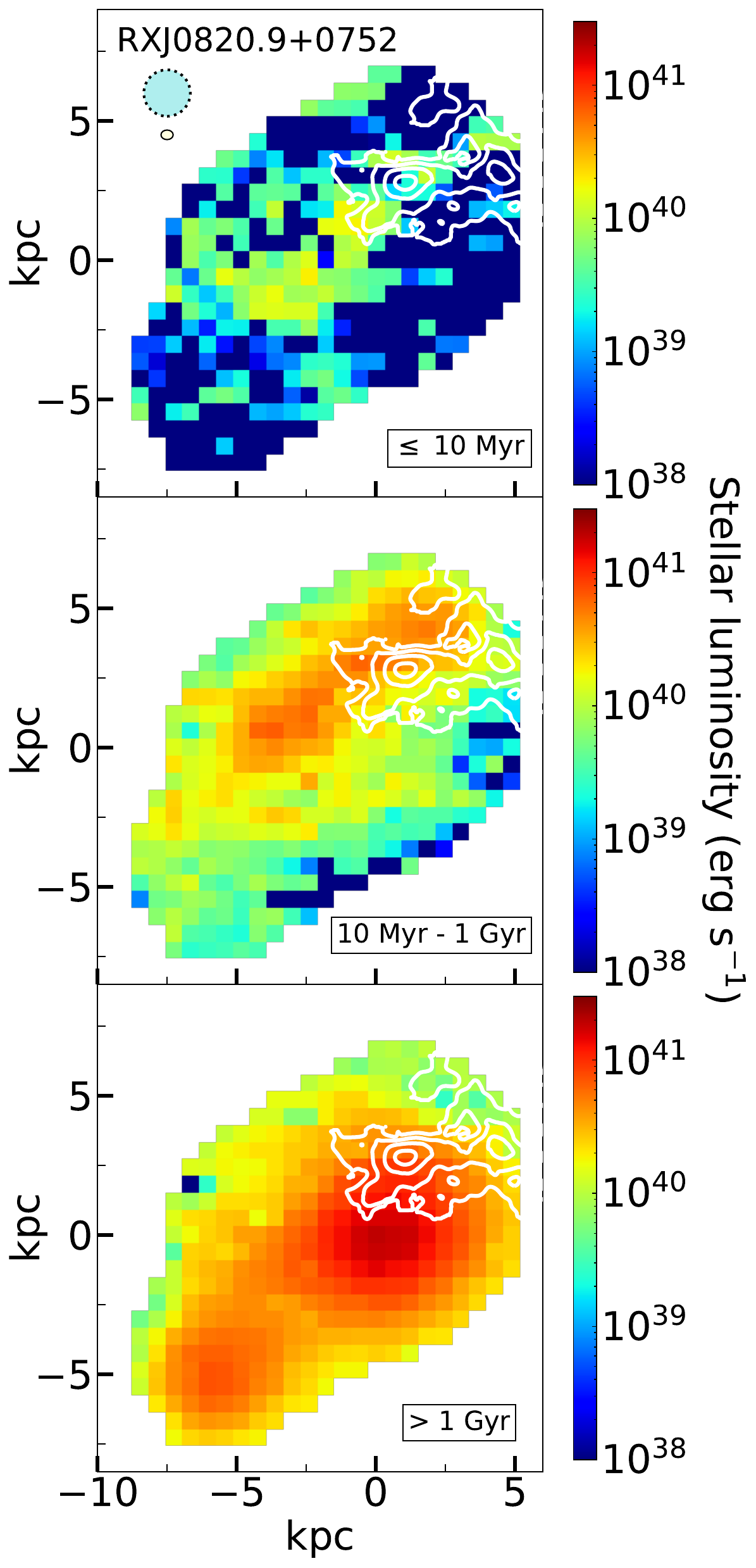}
    \caption{Stellar luminosity distributions in the central galaxy of  RX J0820.9+0752 for young (top panel), intermediate-age (middle panel) and old (bottom panel) stars. The white contours show ALMA CO(3$-$2) emission flux, which represent 10\%, 20\%, 40\%, and 80\% of the maximum CO(3$-$2) flux \citep{Vantyghem2017}. The solid line ellipse shows the beam size of the ALMA CO(3$-$2) observations and the dashed circle shows the seeing of our KCWI observations.
    }
    \label{fig:RXJ0820:Luminosities_ages}
\end{figure}

\begin{deluxetable*}{ccccccccc}
\tabletypesize{\scriptsize}
\tablewidth{0pt} 
\tablecaption{Properties of Regions of Interest \label{tab:RXJ0820_regions}}
\tablehead{ \colhead{Target} & \colhead{Region} &\colhead{$E(B-V)_{\text{gas}}$} &\colhead{$E(B-V)_{*}$} & \colhead{$D_n$(4000)} & \colhead{Young} & \colhead{intermediate-age} & \colhead{Old} & \colhead{Depletion timescale}\\ \colhead{}  & \colhead{}  &\colhead{(mag)} &\colhead{(mag)} &\colhead{} &\colhead{(\%)} & \colhead{(\%)} & \colhead{(\%)} & \colhead{(Myr)}\\ }
\colnumbers
\startdata 
{} & Main galaxy& $0.2\pm0.2$& 0.00$\pm0.01$ & 2.11 & 0.15 & 8.5 & 91.5 &\\
RX J0820.9$+$0752 & Secondary galaxy& - & 0.12$\pm0.03$ &2.68 & 2 & 8 & 89 & $\sim 7000$\\
{} & Blue arc& $0.46\pm0.09$ & 0.04 & 1.59 & 2 & 52 & 45 &\\
\hline
{} & East&0.3$\pm$0.1 & 0.04$\pm$0.02 &  1.42&45& 34&21 &\\
A1835 &Central&$0.5\pm0.2$  &$0.00\pm0.01$ &1.46&30&53&17 & $\sim 500$\\
{} & West&$0.4\pm0.4$  &$0.00\pm0.03$  &1.54&15&36&48\\
\hline
PKS 0745-191 & Main galaxy&$0.83\pm0.08$ &$0.31\pm0.03$  & 1.65 & 27 & 11 & 62 & $\sim 625$\\
{} & Secondary galaxy& - & $0.23\pm0.05$ & 2.70 & 7 & 0 & 93 &\\
\enddata
\tablecomments{(3) Nebular broadband color excess obtained using the Case B extinction curve from \cite{Cardelli1989} with $R_V$=3.1 and the ratio of $H{\beta}$/$H{\gamma}$. (4) Stellar broadband color excess obtained from continuum fits using the extinction curve from \cite{Calzetti2000} using $R_V$=4.05. Uncertainties show the 1$\sigma$ errors obtained from 100 Monte Carlo iterations. (5) Balmer break index \citep{Balogh1999}, which is the ratio of the blue continuum band flux (3750-3850 \AA) to the red continuum band flux (4150-4250 \AA): $\int_{3750}^{3850}F_{\nu}(\lambda) d\lambda / \int_{4150}^{4250}F_{\nu}(\lambda) d\lambda$. These bands were chosen for their lack of emission lines in our three targets. (6)$-$(8) Relative contribution of young ($\leq 10$ Myr), intermediate-age (10 Myr$-$1Gyr) and old ($> 1$ Gyr) stars to the stellar continuum. The distribution of stellar ages within a region is obtained from the relative contribution of each SPS template to the starlight fit at a rest-frame wavelength of 5000 \AA (see \autoref{sec:Methods:Stellar_fit}). (9) Depletion timescale: Time for star formation to deplete the current molecular gas reservoir in each galaxy. For RX J0820.9+0752, the average SFR over the past 500 Myr is used. For A1835 and PKS 0745-191, the average SFR over the past 10 Myr is used.}
\end{deluxetable*}

The center of RX J0820.9+0752 is highly asymmetric, which has be seen across multiple wavelength bands \citep{Bayer-Kim2002,Vantyghem2019}. Its nebular and molecular gas lie $\sim$20 kpc northwest of the peak stellar emission \citep{Bayer-Kim2002,Salome&Combes2004,Vantyghem2017,Vantyghem2019,Olivares2019}. RX J0820.9+0752 is also known for its massive $10^{10}$ M$_{\odot}$ reservoir of cold molecular gas \citep{Vantyghem2017} and for its peculiar morphology, with two blue clumpy arcs northeast of the central galaxy with strong and dusty line emission \citep{Bayer-Kim2002}.

From top to bottom, \autoref{fig:RXJ0820:flux;vstel;v50-vstel} shows the stellar continuum flux map, stellar median velocity map ($v_*$) and nebular gas velocity with respect to the stars ($v_{50}^{\text{[OII]}}-v_*$). We detect stellar emission in a region spanning roughly 14 kpc across. The total stellar luminosity is $2.3\times 10^{43}$ erg s$^{-1}$ between observed wavelengths of $3700-5650$ \AA.

The top panel of \autoref{fig:RXJ0820:flux;vstel;v50-vstel} shows two distinct peaks corresponding to the two galaxies within our FOV: the main/brightest galaxy, taken to be the center of the system, and a secondary galaxy located $\sim$ 7.8 kpc southeast. 

In the middle panel of \autoref{fig:RXJ0820:flux;vstel;v50-vstel} positive velocities correspond to redshifted stars (with respect to the systemic velocity) and negative velocities are blueshifted. The solid grey lines outline the continuum flux, with contours showing 20\%, 40\% and 80\% of the peak continuum flux. The velocities of the stars in the nucleus are close to 0, by definition. The  continuum flux-weighted mean stellar velocity is $\sim 50$ km s$^{-1}$. We use flux-weighted velocities to estimate the average velocity of stars or gas  within a region. We define this as:
\begin{equation}\label{eq:flux-weighted_mean}
    \bar{v}=\frac{\sum_{i} v_i \times f_i}{\sum_i f_i}
\end{equation}
where $v$ is the velocity and $f_i$ is the flux within the i{\it th} spaxel of any given region.
The stellar velocity map shows two redshifted regions, one north and one south, with $v_* \sim$$+175$ km s$^{-1}$ and $v_* \sim$$+130$ km s$^{-1}$, respectively. The southern redshifted region coincides with the position of the secondary galaxy. Two slightly blueshifted regions lie to the east and west of the nucleus.

The bottom panel of \autoref{fig:RXJ0820:flux;vstel;v50-vstel} compares the median nebular gas velocity to the velocity of the stars in each spaxel. The median velocity of the [OII] doublet is used for the nebular gas velocity as it is the brightest emission line for our three targets. \footnote{See \cite{Gingras2024} for more explanation on the emission line fitting methodology.} The dashed dotted ellipse shows the position of a low significance X-ray cavity \citep{Vantyghem2019}. 

The $v_{50}^{\text{[OII]}}-v_*$ map shows little blueshifted gas with respect to the stars. \cite{Gingras2024} detected a bulk motion between the nebular gas and the central stars of $+150$ km s$^{-1}$ in RX J0820.9+0752, which is seen by the overall redshifting of the nebular gas in the bottom panel of \autoref{fig:RXJ0820:flux;vstel;v50-vstel}.
The largest velocity difference between the nebular gas and the stars is seen west of the nucleus, where $v_{50}^{\text{[OII]}}-v_* \sim +175-300$ km s$^{-1}$. The stars in this region are slightly blueshifted by $\sim -35$ km s$^{-1}$ (see central panel), whereas the nebular gas is redshifted. North of this region, the stars have similar velocities to the nebular gas, with $v_{50}^{\text{[OII]}}-v_* \sim +40$ km s$^{-1}$.

\autoref{fig:regions:RXJ0820} focuses on three regions of interest within our FOV: the main galaxy, the secondary galaxy, and the blue continuum arc identified by \cite{Bayer-Kim2002}. A Hubble Space Telescope (HST) image of the blue arc is shown in the upper right of \autoref{fig:regions:RXJ0820} \citep{Bayer-Kim2002,Vantyghem2017}. The grey region in \autoref{fig:regions:RXJ0820} shows the stellar continuum footprint while the colored regions show the spaxels used when making spatially-integrated spectra of the three regions of interest. 

\autoref{tab:RXJ0820_regions} lists various properties of the regions shown in \autoref{fig:regions:RXJ0820}, including their broadband color excess from both nebular and stellar emission, the Balmer break index (D4000), and the relative contribution of young ($\leq 10$ Myr), intermediate-age (10 Myr$-$1Gyr) and old ($> 1$ Gyr) stars to the stellar continuum flux. The uncertainties on $E(B-V)$ values quoted in \autoref{tab:RXJ0820_regions} reflect only the statistical errors from the fit. The true uncertainties are likely larger due to systematic effects. \autoref{tab:RXJ0820_regions} also lists the depletion timescale for each system, which is the time required to exhaust the galaxy's molecular gas reservoir based on its star formation rate.

The main galaxy has little dust, with $E(B-V)\sim 0$ both for the gas and for the stars. This is also the case for the secondary galaxy, with $E(B-V)_* = 0.12 \pm 0.03$. Following \cite{Martis2016}, we consider a region to be dusty if $E(B-V) \geq 0.3$. On the other hand, the nebular gas associated with the blue continuum arc is dusty, with $E(B-V)_{\textrm{gas}} = 0.46$. This is consistent with observations by \cite{Bayer-Kim2002} which measure $E(B-V)$ between 0.3$-$0.7 in the various star formation knots of the blue arcs.

\autoref{fig:RXJ0820:Ages_maps} shows the spatial distributions of the fractions of young, intermediate-age and old stars. These fractions are determined by measuring the contribution of the SPS models in each age bin to the fitted continuum. Similarly, \autoref{fig:RXJ0820:Luminosities_ages} maps the luminosity from stars of different ages. The white contours show the ALMA CO(3$-$2) emission in the core of  RX J0820.9+0752 \citep{Vantyghem2017}.

The spectra in \autoref{fig:regions:RXJ0820} show that the stellar populations in the main and projected galaxies are old. They have strong Balmer breaks and lack Balmer absorption lines. This is also seen in \autoref{fig:RXJ0820:Ages_maps} and \autoref{fig:RXJ0820:Luminosities_ages} where almost all of the stellar emission, 92\% for the main galaxy and 89\% for the secondary galaxy, is emitted by stars older than 1 Gyr. 

On the other hand, the blue arc has a significant intermediate-age stellar population, with half of the starlight coming from intermediate-age stars and half from old stars (presumably background stars). This is confirmed by the spatially-integrated spectrum of the blue arc in \autoref{fig:regions:RXJ0820}, which shows strong Balmer absorption lines. As the main and secondary galaxies are dominated by older stars, it is unlikely that the gas and young stars in the blue arc were stripped from either galaxy. The stars likely formed out of the surrounding gas, with gravitational interactions between the two galaxies potentially triggering the cooling process. The bottom panel of \autoref{fig:RXJ0820:flux;vstel;v50-vstel} shows similar velocities between the stars and the nebular gas in the blue arc, which further supports this scenario.

The top panels of \autoref{fig:RXJ0820:Ages_maps} and \autoref{fig:RXJ0820:Luminosities_ages} show a dearth of stars younger than 10 Myr throughout the center of RX J0820.9+0752. Only $\sim 2\%$ of the starlight in the blue arc comes from younger stars. This indicates that the strong line emission in the blue arc is unlikely to have been caused by photoionization from young O and early B type stars. These massive stars have short lifetimes ($\lesssim10–20$ Myr) and dominate the production of ionizing photons. In the absence of a substantial population younger than 10 Myr, the ionizing photon production rate declines rapidly \citep{Leitherer1999}. Even if the intermediate-age stars observed in the blue arc were on the younger side of the $10$ Myr $-$ $1$ Gyr age range, the amount of ionizing photons emitted would be insufficient to explain the nebular emission observed. A different ionization mechanism, such as shocks or cosmic ray heating \citep{Pfrommer2008,Ruszkowski2008}, is most likely responsible for the widespread nebular emission in RX J0820.9+0752.

Despite its lack of young stars, the blue arc hosts a substantial intermediate-age stellar population, with a mass of almost $7\times10^8$ M$_{\odot}$, corresponding to an average star formation rate of 1.4 M$_{\odot}$ yr$^{-1}$ over the past 500 Myr. RX J0820.9+0752's $10^{10}$ M$_{\odot}$ reservoir of molecular gas is a potential fuel source for these stars as it partially overlap with the blue arc \citep{Vantyghem2017} (see the central panel of \autoref{fig:RXJ0820:Luminosities_ages}). If this star formation had continued at a rate of $1.4$ M$_{\odot}$ yr$^{-1}$, it would have had ample molecular gas to consume, with a depletion timescale of $\sim 7$ Gyr.


\subsection{A1835}\label{sec:Results:Abell 1835}

\begin{figure}
    \centering
    \includegraphics[width=\columnwidth]{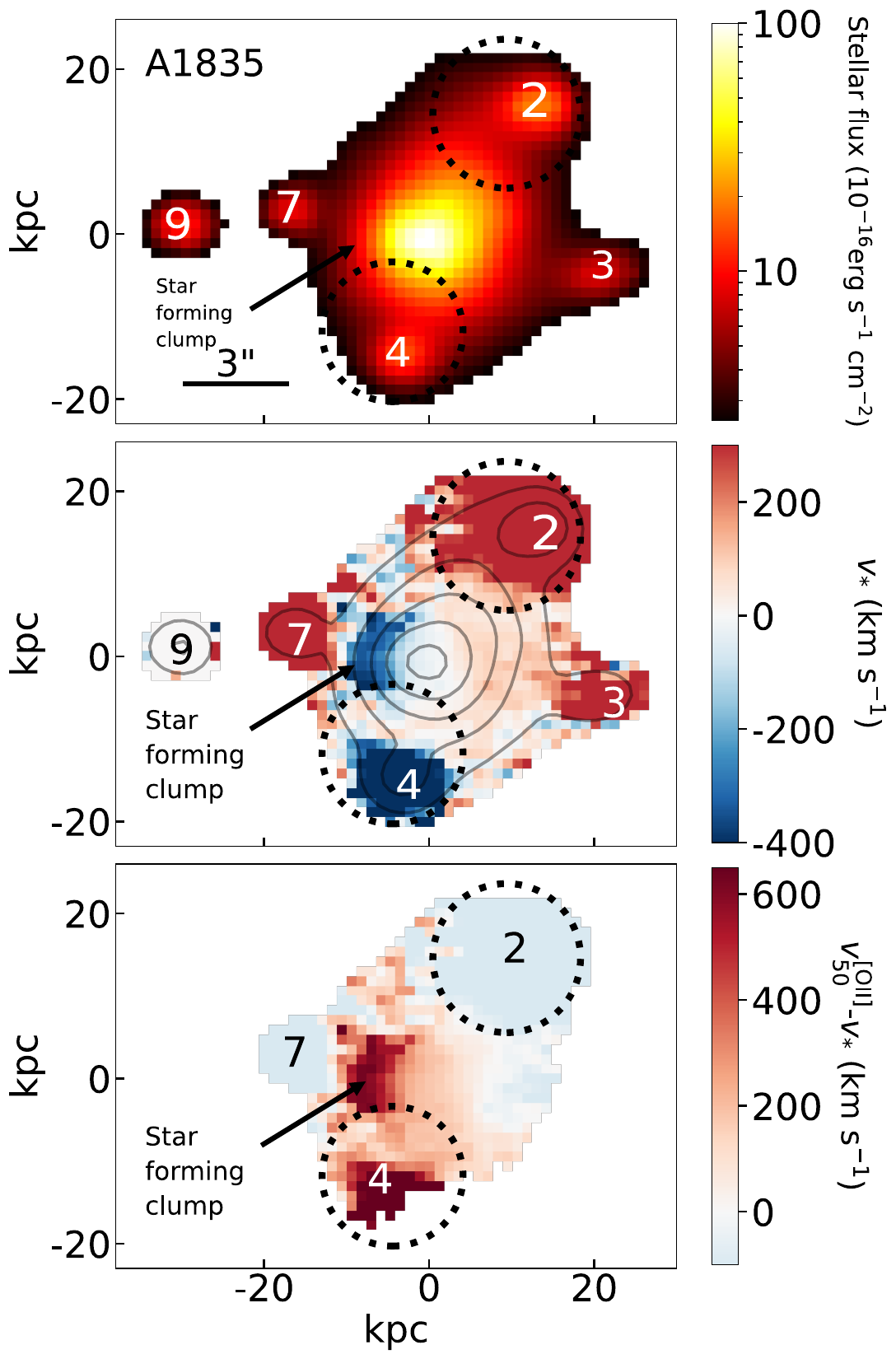}
    \caption{Stellar properties maps for the central galaxy of A1835. The black dotted ellipses show the positions of X-ray cavities \citep{McNamara2006,Olivares2019,Russell2019}. The numbers identify the projected galaxies, labeled as in \autoref{fig:A1835_galaxies_spectrum} in \autoref{appendix:A1835_galaxies}. Top: Stellar continuum flux map for rest-frame wavelengths of $3435-5035$ \AA. Only spaxels with stellar fluxes of at least $2.5 \times 10^{-16}$ erg s$^{-1}$ cm$^{-2}$ are included. Middle: Median stellar velocity map where grey contours show stellar fluxes of 5\%, 10\%, 20\%, 40\% and 80\% of the maximum stellar flux per spaxel. Bottom: Velocity of the nebular gas with respect to the stars.}
    \label{fig:A1835:flux;vstel;v50-vstel}
\end{figure}

A1835 is a strong cooling core cluster undergoing powerful radio-mechanical AGN feedback \citep{McNamara2014}. Its central galaxy has a SFR of $\sim 100-180$ M$_{\odot}$ yr$^{-1}$ in the inner $\sim 35$ kpc and contains $\sim 5\times 10^{10}$ M$_{\odot}$ of molecular gas in the central $\sim 10$ kpc \citep{McNamara2006,McNamara2014}.

Analogous to \autoref{fig:RXJ0820:flux;vstel;v50-vstel}, \autoref{fig:A1835:flux;vstel;v50-vstel} maps the stellar continuum flux, stellar velocity ($v_*$), and nebular gas velocity with respect to the stars ($v_{50}^{\text{[OII]}}-v_*$). The numbers show the positions of projected galaxies as labeled in \autoref{fig:A1835_galaxies_spectrum} in \autoref{appendix:A1835_galaxies} and the dashed circles show the positions of the X-ray cavities \citep{McNamara2006,Olivares2019,Russell2019}. The top panel shows the stellar flux in the inner region; the larger FOV, which includes eight projected galaxies, is shown in \autoref{appendix:A1835_galaxies}. There, we show that none of the projected galaxies appear to be impacting the dynamics of the central galaxy. Therefore, \autoref{fig:A1835:flux;vstel;v50-vstel} only includes the central region where nebular emission is observed. Five projected galaxies are present in the central stellar emission region shown in the top panel of \autoref{fig:A1835:flux;vstel;v50-vstel}. The total stellar luminosity shown in the top panel of \autoref{fig:A1835:flux;vstel;v50-vstel} between observed wavelengths of $4300-6300$ \AA{} is $1.9\times 10^{44}$ erg s$^{-1}$.

The central panel of \autoref{fig:A1835:flux;vstel;v50-vstel} shows that the stellar velocities in the core of A1835 lie between $\sim -400$ km s$^{-1}$ and $\sim +200$ km s$^{-1}$, with velocities increasing (more redshifted) from east to west. Although there is a stellar velocity gradient from east to west, the stellar velocities are not consistent with rotation, since the blueshifted and redshifted velocities differ in magnitudes by hundreds of km s$^{-1}$.
The bottom panel of \autoref{fig:A1835:flux;vstel;v50-vstel} shows that most of the nebular gas is redshifted with respect to the stars, with an overall redshift of $\sim 150$ km s$^{-1}$ \citep{Gingras2024}. The main feature in this velocity map is the region east of the nucleus where the velocity difference between the nebular gas and the stars is $\sim + 600$ km s$^{-1}$. This corresponds to the highlighted region in the stellar velocity map, where the stars are blueshifted by $\sim 300$ km s$^{-1}$. 

\begin{figure}
    \centering
    \includegraphics[width=\columnwidth]{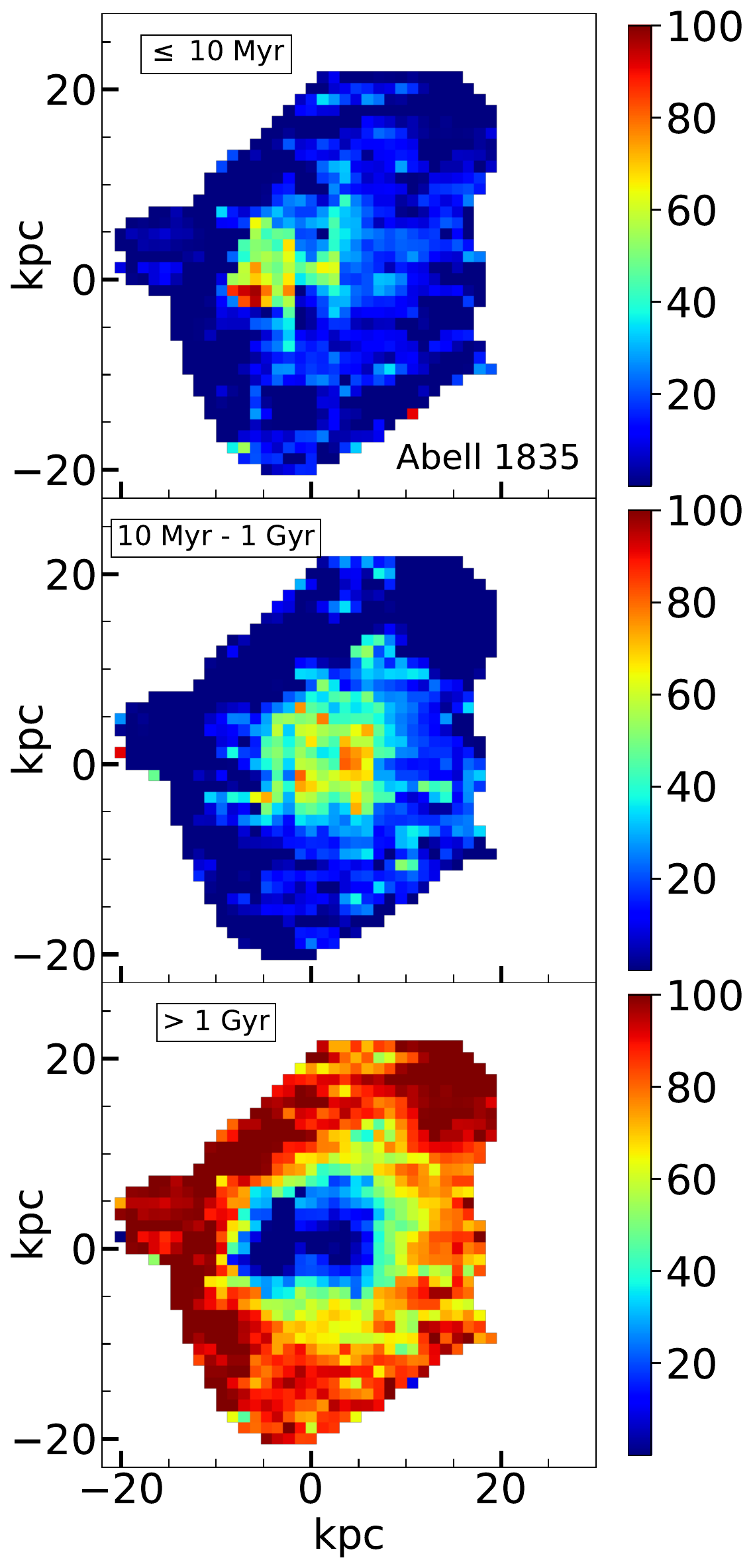}
    \caption{Distributions of stellar ages in the central galaxy of A1835. The colorbars represent the fraction of the stellar flux emitted by young ($\leq$ 10 Myr), intermediate-age (10 Myr $-$ 1 Gyr) and old ($>$ 1 Gyr) stars.}
    \label{fig:A1835:Ages_maps}
\end{figure}

\begin{figure}
    \centering
    \includegraphics[width=\columnwidth]{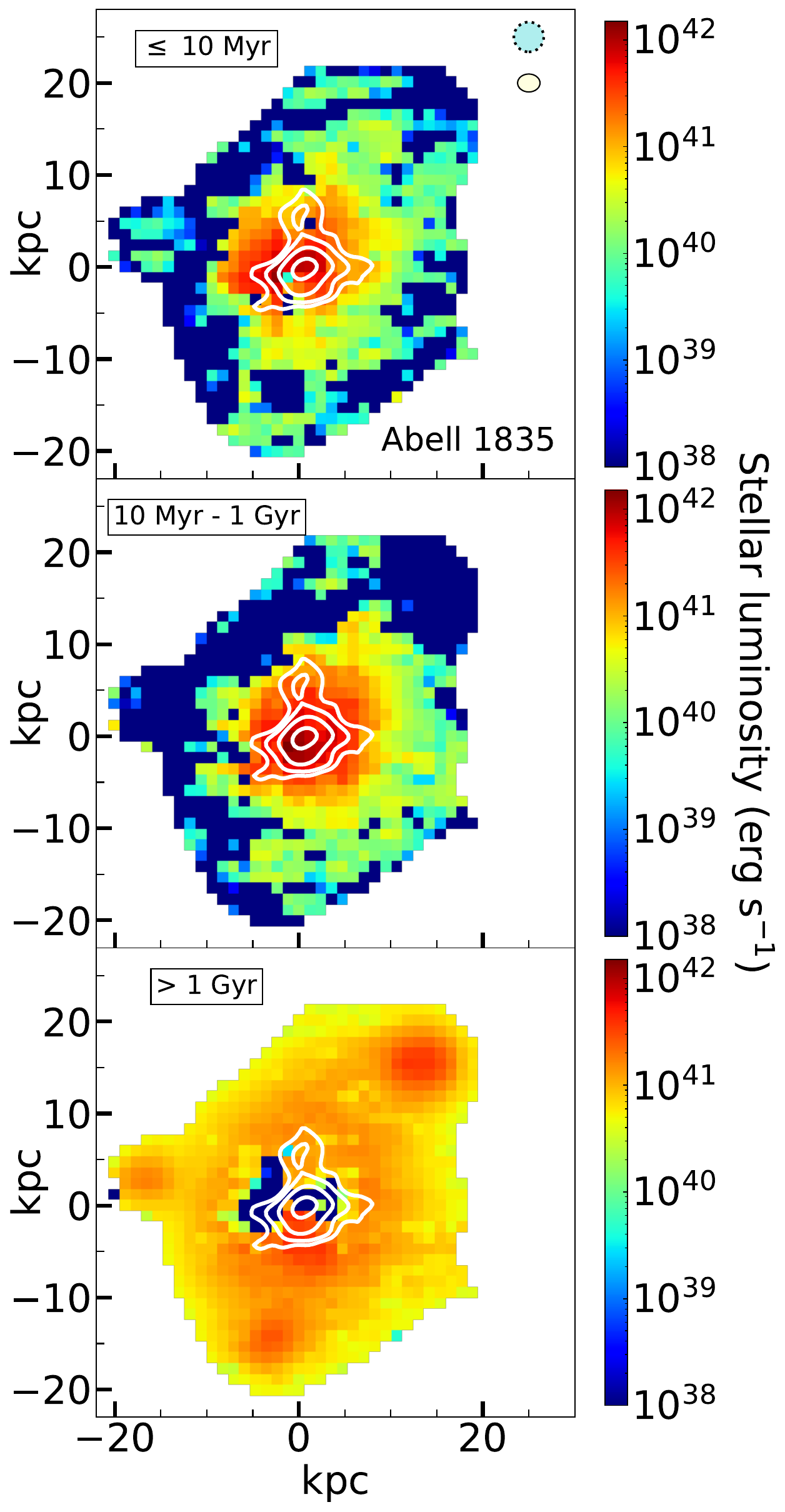}
    \caption{Stellar luminosity distributions for young (top panel), intermediate-age (middle panel) and old (bottom panel) stars in the central galaxy of A1835. The white contours show ALMA CO(3$-$2) emission flux, which represent 10\%, 20\%, 40\%, and 80\% of the maximum CO(3$-$2) flux \citep{McNamara2014}. The solid line ellipse shows the beam size of the ALMA CO(3$-$2) observations and the dashed circle shows the seeing of the KCWI observations.
    }
    \label{fig:A1835:Luminosities_ages}
\end{figure}

\autoref{fig:A1835:Ages_maps} shows the spatial distributions of the relative fractions of young, intermediate-age, and old stars in the core of A1835. Similarly, \autoref{fig:A1835:Luminosities_ages} maps the luminosities of young, intermediate-age, and old stars. Both figures show extended young and intermediate-age stellar populations spanning a radius of $15-20$ kpc around the nucleus. The luminosity maps further show that the young and intermediate-age stars spatially overlap with the detected molecular gas. If the more extended distributions of young and intermediate-age stars trace the full molecular gas distribution, this would imply that additional molecular gas may exist below our detection limit. We reached a similar conclusion in our previous paper using the distribution of [OII]3726,9 \AA{} flux, inferring a possible $M_{\rm{mol}}$ of more than $10^{11}$ M$_{\odot}$ \citep{Gingras2024}.

\begin{figure*}
    \centering
    \includegraphics[width=\textwidth]{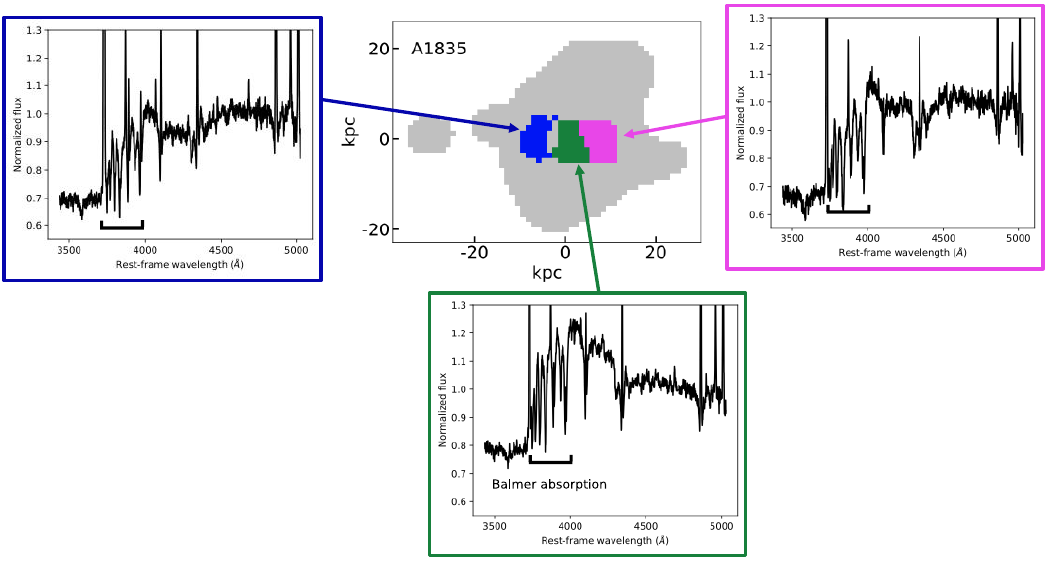}
    \caption{Spectra of three central regions in A1835. The grey region shows the footprint of the stellar emission as in the top panel of \autoref{fig:A1835:flux;vstel;v50-vstel}. }
    \label{fig:regions:A1835}
\end{figure*}

\autoref{fig:regions:A1835} separates the core of A1835 in three regions of interests and shows their spatially-integrated spectra. These three regions were selected using the stellar velocity map in the central panel of \autoref{fig:A1835:flux;vstel;v50-vstel}: eastern blueshifted region (blue), central region (green), and western redshifted region (magenta). \autoref{tab:RXJ0820_regions} lists their properties. The nebular emission in the center of A1835 is dusty, with all three central regions having $E(B-V)_{\text{gas}} \geq 0.3$. These values are consistent within $1\sigma$ with those listed in \cite{Crawford1999}. 

The extended intermediate-age stellar population in the core of A1835 is clearly seen in \autoref{fig:regions:A1835}, where deep Balmer absorption lines characteristic of A-type stars are present in all three spectra. This is also reflected in their $D_n$(4000) values listed in \autoref{tab:RXJ0820_regions}. All three regions therefore contain substantial young and intermediate-age stellar populations, with more than 50\% of the stellar flux arising from stars younger than 1 Gyr. \autoref{fig:A1835:Ages_maps} and \autoref{tab:RXJ0820_regions} reveal differences in the stellar age distributions across the core of A1835. Both the Balmer break indices and the fractions of young, intermediate-age, and old stars show a clear east–west gradient in stellar ages.

Overall, the coexistence of young ($<10$ Myr) and intermediate-age (10 Myr$–$1 Gyr) stars indicates that star formation has been sustained over an extended period, rather than triggered only recently. If star formation had begun only in the past few Myr, the stellar light would be dominated by very young massive stars, and a substantial intermediate-age population would not yet have had time to develop. The observed stellar populations therefore imply long-lived star formation in the central $15–20$ kpc.

Throughout our FOV, we estimate an average SFR of $40$ M$_{\odot}$ yr$^{-1}$ for the past Gyr, with an increased SFR of $100$ M$_{\odot}$ yr$^{-1}$ in the past 10 Myr. These star formation rates are consistent with those measured by \cite{Crawford1999} and \cite{McNamara2006}. As discussed in \autoref{appendix:A1835_galaxies}, all the projected galaxies have old stellar populations, making it unlikely that stripped gas from these projected galaxies fueled this star formation. Therefore the galaxy's massive reservoir of molecular gas and the starbust it fuels likely cooled from the surrounding hot atmosphere. Fueled by $5\times10^{10}$ M$_{\odot}$ of molecular gas \citep{McNamara2014}, a SFR of 100 M$_{\odot}$ yr$^{-1}$ implies a depletion timescale of $\sim500$ Myr. In order to sustain the current SFR of 100 M$_{\odot}$ yr$^{-1}$, the reservoir of molecular gas fueling this star formation will need to be replenished by newly-formed cool gas.

\subsubsection{Young Stars Forming in an Outflow?}\label{sec:A1835:clump}

\begin{figure*}
    \centering
    \includegraphics[width=\textwidth]{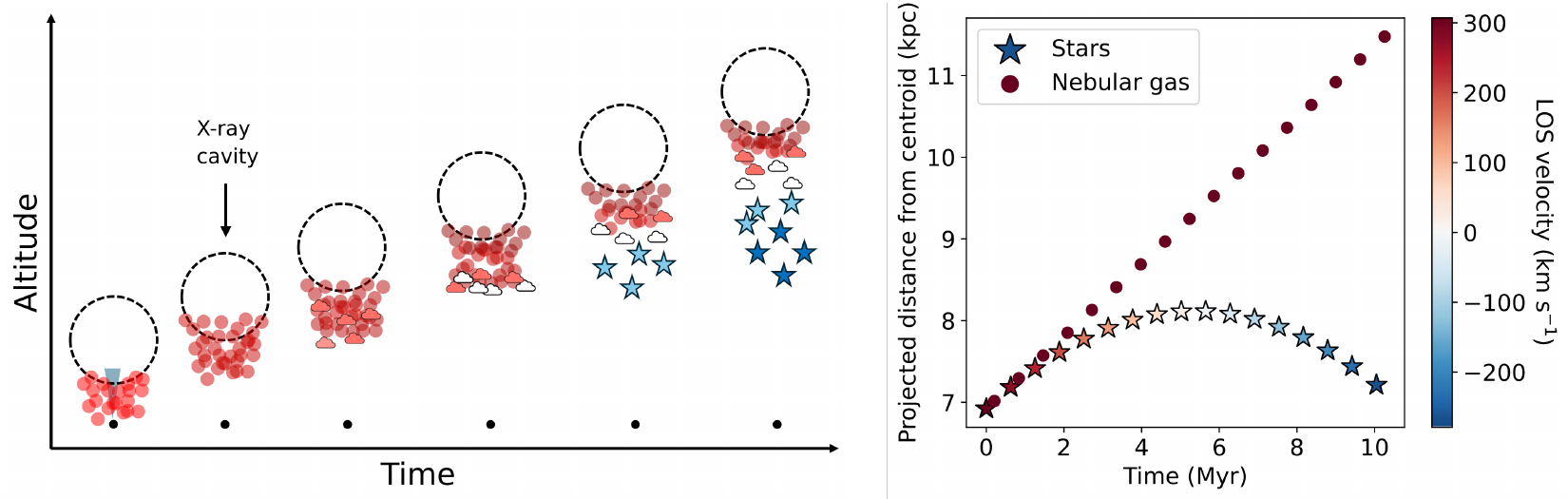}
    \caption{Left: Diagram of the young stars forming in an outflow scenario that is discussed in \autoref{sec:A1835:clump}. The change of colors, from red to blue, denotes the change in line-of-sight velocity from redshifted to blueshifted. Right: Time evolution of the projected distance from the central AGN to the outflowing nebular gas and to the star-forming clump discussed in \autoref{sec:A1835:clump}. The colorbar shows the change in the line-of-sight velocity of the newly-formed stars.}
    \label{fig:A1835:Diagram}
\end{figure*}

\autoref{fig:regions:A1835} shows a large clump of star formation $5-10$ kpc east of the nucleus (blue region). This accounts for $7-10\%$ of the observed starlight (shown in the top panel of \autoref{fig:A1835:flux;vstel;v50-vstel}). Its spectrum, in \autoref{fig:regions:A1835}, shows strong nebular line emission and strong Balmer absorption features consistent with a young but aging population. Its stellar mass for stars younger than 10 Myr is $1.65\times10^8$ M$_{\odot}$, which corresponds to an average SFR of 16.5 M$_\odot~\text{yr}^{-1}$ over the past 10 Myr. It has an average SFR over the past Gyr of 3.5 M$_\odot~\text{yr}^{-1}$, showing a recent increase in the region's star formation.

The bottom panel of \autoref{fig:A1835:flux;vstel;v50-vstel} shows that the velocities of the stars and the nebular gas in this region, projected along the same line-of-sight, are offset by $\simeq 600$ km s$^{-1}$. The flux-weighted stellar velocity is blueshifted with respect to the central galaxy by $-280$ km s$^{-1}$ while the nebular gas is redshifted by $+ 307$ km s$^{-1}$. CO(1$-$0) emission is detected in this region, indicating a molecular gas mass of $5\times 10^9 - 1\times10^{10}$ M$_{\odot}$ close to this line-of-sight. This molecular gas, which is shown in \autoref{fig:A1835:Luminosities_ages} \citep{McNamara2014}, is redshifted with respect to the galaxy by $\sim 500~\text{km}~\text{s}^{-1}$, receding slightly faster than the nebular gas.
These large velocity differences would suggest at first glance that the gas and stars along the line-of-sight are unrelated. However, it would be surprising if such a large, coherent region of recent star formation formed with no surrounding gas to fuel it.  We therefore explore the possibility that the two are related, but that the stars have decoupled kinematically from their birth clouds. 

The high relative speeds of the system could be explained naturally by tidal or ram-pressure stripping from a flyby galaxy or as debris in a late-stage merger. While these possibilities cannot be excluded outright, they present several issues. Firstly, none of the eight galaxies projected along the line-of-sight within 60 kpc of the central galaxy contain emission lines or blue populations capable of donating upward of $10^{10}~\rm M_\odot$ of molecular gas, nor are there clear signs of tidal tails and debris linked to these galaxies. All appear to be old red ellipticals.

A late-stage merger is another possibility. However a galaxy, likely a spiral, containing so much molecular gas would be rare in a cluster at this redshift. Finding its way into the center of the cluster where few star-forming galaxies reside would be equally unlikely \citep{Dressler1980,Dressler1997,Balogh2004,Boselli2006}.

The enormous amount of molecular gas in the core of A1835 is believed to have originated from in-situ cooling \citep{McNamara2014,Gingras2024}. In this context we consider how such a high-speed population would form. Gas condensing into stars at large radii and falling inward is a possibility. The velocity differences between the stars and gases could arise as the stars decoupled from the gas and began falling ballistically with the trailing gases descending more slowly due to ram pressure drag. This scenario is inconsistent with our observations, however. For stars to form out of the cooling gas they must initially lie on the same side of the galaxy, either both in front of or both behind the nucleus. If both are infalling their line-of-sight velocities should have the same sign. However, we observe the young stars to be blueshifted while the nebular gas along the same line-of-sight is redshifted. Therefore, if the stars formed out of the gas, they must be moving in opposite directions, which rules out this simple infall scenario.


We instead focus on the more likely possibility that the stars formed in a nebular gas outflow driven by the central radio jets and X-ray bubbles. \cite{Gingras2024} showed that nebular gas velocities in this region are consistent with a bipolar outflow 
in the same direction as the clumps of gas and star formation. The young stars may have formed in this outflow over the past 10 Myr but have since decoupled from their birth clouds.

Such a scenario has been previously suggested, based on both simulations and observations. Significant spatial and/or velocity offsets between young stars and multiphase gas have been reported in other systems, such as A1795 \citep{Tamhane2023}, SDSS J1531+3414 \citep{Omoruyi2024}, IRAS F23128$-$5919 \citep{Maiolino2017}, and NGC 1275 \citep{Canning2010,Canning2014}. For these systems, projected separations of $\sim0.6–3$ kpc (where spatially resolved) and line-of-sight velocity differences of $\sim150–800$ km s$^{-1}$ have been measured. These offsets have been suggested to arise from young stars forming out of cooling multiphase gas and subsequently decoupling from it. This has also been observed in simulations, where AGN activity, jets and/or X-ray cavities uplift low-entropy gas to higher altitudes, some of which then cools, forms stars, and falls back towards the central AGN \citep{Revaz2008,Li2015}.

Such a scenario is shown in the left panel of \autoref{fig:A1835:Diagram} for the blue star-forming clump in A1835. It shows the time evolution of outflowing nebular gas as it is driven in the wake of an X-ray cavity, cooling and ultimately forming stars. While X-ray cavities may be uplifting the outflowing gas, we have yet to identify the mechanism responsible for driving the gaseous outflow. The nebular gas outflow, which spatially coincides with the young star clump, does not have a detected X-ray cavity associated with it. However, there are numerous reasons why X-ray cavities may not be detected. They can become difficult to detect due to buoyant rise, deformation and mass entrainment which reduce the X-ray contrast, projection effects, and low signal-to-noise \citep{Ensslin2002,Birzan2004,Bruggen2009}.

If we assume that the forming stars are initially moving with the nebular gas outflow at a line-of-sight velocity of $ +307~\text{km}~\text{s}^{-1}$, we can calculate the timescale on which they will reverse their motion and ultimately fall ballistically towards the central AGN at velocities of $- 280$ km s$^{-1}$. \cite{Hogan2017} and \cite{Pulido2018} modeled the gravitational potential of A1835 as a combination of an isothermal and NFW (Navarro–Frenk–White) potential and inferred a total mass of $\sim 9 \times 10^{11}$ M$_\odot$ within a radius of 7 kpc from the Chandra X-ray centroid. \cite{Hogan2017_b} shows that this gravitational potential model gives reliable total enclosed masses at the radius studied here. Knowing the total enclosed mass and the distance from the center, we calculate a gravitational acceleration at 7 kpc of $-2.5\times10^{-12}$ km s$^{-2}$, which is used throughout these calculations. The right panel of \autoref{fig:A1835:Diagram} shows how the altitude and the line-of-sight velocity of the nebular gas and the most recently formed stars evolve with time. The forming/newly-formed stars are assumed to be falling ballistically without drag, from an initial line-of-sight velocity of $+307$ km s$^{-1}$ to a final observed velocity of $-280$ km s$^{-1}$. Therefore, knowing their initial and final velocities, as well as the gravitational potential \citep{Hogan2017,Pulido2018}, we can calculate the star-forming clump's position and velocity throughout this process, as shown in \autoref{fig:A1835:Diagram}. It takes roughly 10 Myr for the newly-formed stars to decouple from the nebular gas outflow and infall towards the galaxy center at a line-of-sight velocity of $-280$ km s$^{-1}$. The dark red dotted line in \autoref{fig:A1835:Diagram} shows the motion of the outflowing nebular gas, which is still being uplifted. The line-of-sight velocity of the nebular gas is taken to be constant at $+307$ km s$^{-1}$ \citep[see Figure 15 of][]{Gingras2024}.

Throughout this analysis, we assume that the ouflowing/inflowing nebular gas and stars are moving at a $45^{\circ}$ angle with respect to our line-of-sight, such that their velocities are $\sqrt{2} v_{\text{LOS}}$. The age of the young stars is similar to the time since the decoupling began, implying they could have formed out of the gaseous outflow, decoupled from it, and be now raining back towards the central AGN.

The blueshifted stellar velocities imply that the stars falling inward began their descent on the far side of the galaxy. 
The positive velocities of the nebular and molecular gas indicate they continue to be driven, moving outward on the far side of the galaxy. Their velocity difference indicates that they as well may have separated due to differential ram pressure forces or different coupling with rising bubbles or jet.

A similar mechanism was cited to explain the outflow population in the Seyfert galaxy IRAS F23128$-$5919 \citep{Maiolino2017}.  In that system stars formed roughly 500 pc from the nucleus in a molecular outflow driven by nuclear radiation pressure.  The star formation rate in the outflow population is $\sim 30~\rm M_\odot~yr^{-1}$ representing roughly one quarter of the star formation rate of the entire system. The outflow population and nebular gas velocities differ by $150-350~\rm km~s^{-1}$. Therefore, aside from their different acceleration mechanisms, nuclear radiation pressure in IRAS F23128$-$5919 and, tentatively, cavity uplift in A1835, the two systems are similar. Additionally, in A1835 it is occurring on a somewhat larger scale.

As seen in \autoref{fig:A1835:Diagram}, we expect to see a spatial separation between the nebular gas and the star forming clump as the stars decouple from the outflowing gas. The nebular gas steadily rises away from the central AGN from $\sim 6.9$ kpc to $\sim 11.5$ kpc, while the star-forming clump peaks at a projected distance barely above 8 kpc and then turns around and falls back close to its starting point. This predicts a spatial offset in the sky-plane between the outflowing nebular gas and the now infalling stars of about 4.3 kpc after $\approx 10$ Myr. However, as seen in the bottom panel of \autoref{fig:A1835:flux;vstel;v50-vstel}, the outflowing gas and the blueshifted stars spatially overlap in our observations, forming a region where the velocity difference between the nebular gas and the stars reaches 600 km s$^{-1}$. Since our observations of A1835 have a seeing
of 3.25 kpc and both the outflowing gas and the infalling stars extend over regions larger than the expected spatial offset, we expect the gaseous outflow and the infalling stars to spatially overlap along our line-of-sight. Follow-up observations with the Hubble Space telescope or the James Webb Space Telescope should be able to detect this spatial offset if present. Overall, the observed distributions and kinematics of the young stars and nebular gas in this region are consistent with a scenario where infalling stars formed out of a nebular gas outflow driven by the central AGN.

A similar scenario may be occurring in the Perseus cluster's nebula centred on NGC 1275. \citet{Canning2010,Canning2014} observed that the young blue stars are spatially offset from their parent filaments by $\sim 0.5 - 3.5$ kpc in the outskirts of the nebula. In this instance many are oriented close to the sky plane allowing for clear spatial segregation.

While we cannot exclude other explanations with certainty, the scenario proposed here has precedent in other systems and in simulations. \citet{Revaz2008} and \citet{Gaspari2012} proposed that low entropy gas with masses far exceeding A1835's outflow population can be lifted in the bubble updrafts. This process may accelerate cooling by displacing low-entropy gas to higher altitudes which can trigger thermal instabilities, explaining the filaments of cold gas and star formation often observed in cooling flows \citep{McNamara2016}. The question, raised earlier by \citet{Maiolino2017}, is how important this process may be in galaxy formation and assembly.  

\subsection{PKS 0745-191}\label{sec:Results:PKS0745}

\begin{figure}
    \centering
    \includegraphics[width=\columnwidth]{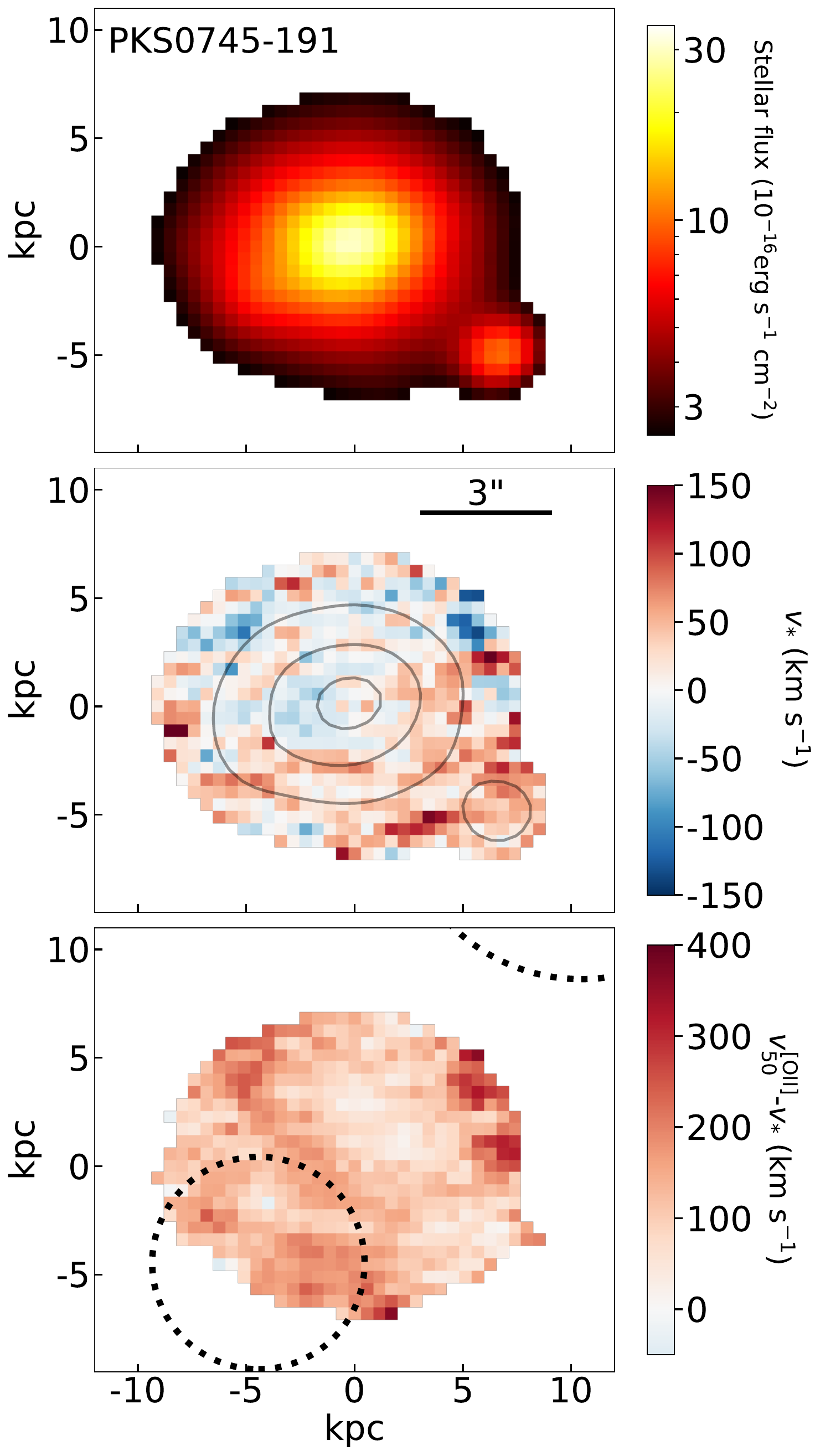}
    \caption{Maps of stellar properties in the central galaxy of PKS 0745$-$191. Top: Integrated stellar continuum flux map (for a wavelength range between $3356-5017$ \AA), with a stellar flux minimum threshold of $2.5 \times 10^{-16}$ erg s$^{-1}$ cm$^{-2}$. Middle: Median stellar velocity map. The grey contours identify 20\%, 40\% and 80\% of the maximum stellar flux per spaxel. Bottom: Median velocity of the nebular gas (from [OII] emission) with respect to the stellar velocity. The black dotted ellipses show the positions of X-ray cavities within our FOV \citep{Sanders2014,Russell2019}.}
    \label{fig:PKS0745:flux;vstel;v50-vstel}
\end{figure}

\begin{figure*}
    \centering
    \includegraphics[width=\textwidth]{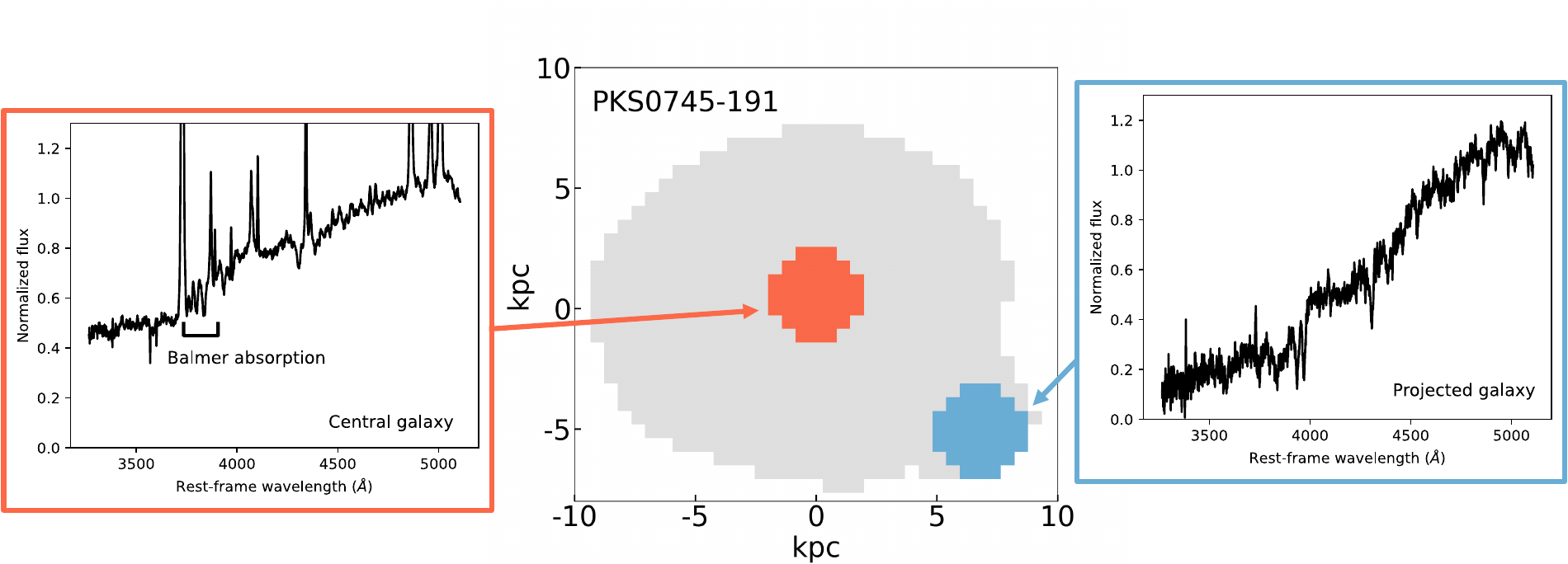}
    \caption{Spectra of the nuclear region of the BCG and the projected galaxy for PKS 0745$-$191: 1\arcsec~around the nucleus of the BCG (red) and 1\arcsec~around the nucleus of the projected galaxy (blue). The grey region shows the footprint of the stellar continuum as seen in the top panel of \autoref{fig:PKS0745:flux;vstel;v50-vstel}. }
    \label{fig:regions:PKS0745}
\end{figure*}

\begin{figure}
    \centering
    \includegraphics[width=\columnwidth]{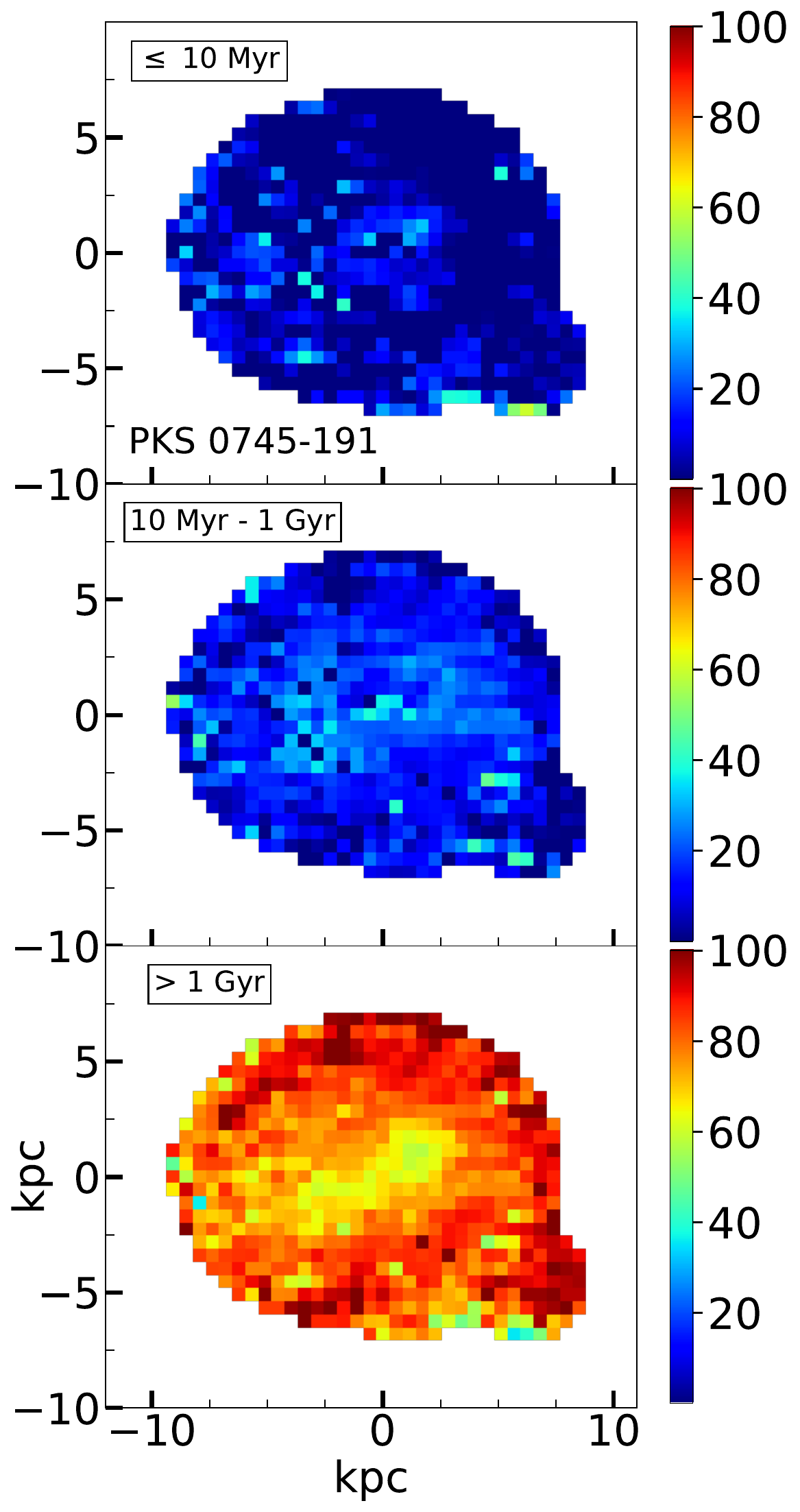}
    \caption{Spatial distributions of stellar ages in the central galaxy of PKS 0745$-$191. The colorbar represents the fraction of stellar flux emitted by young ($\leq$ 10 Myr), intermediate-age (10 Myr $-$ 1 Gyr) and old ($>$ 1 Gyr) stars.}
    \label{fig:PKS0745:Ages_maps}
\end{figure}

\begin{figure}
    \centering
    \includegraphics[width=\columnwidth]{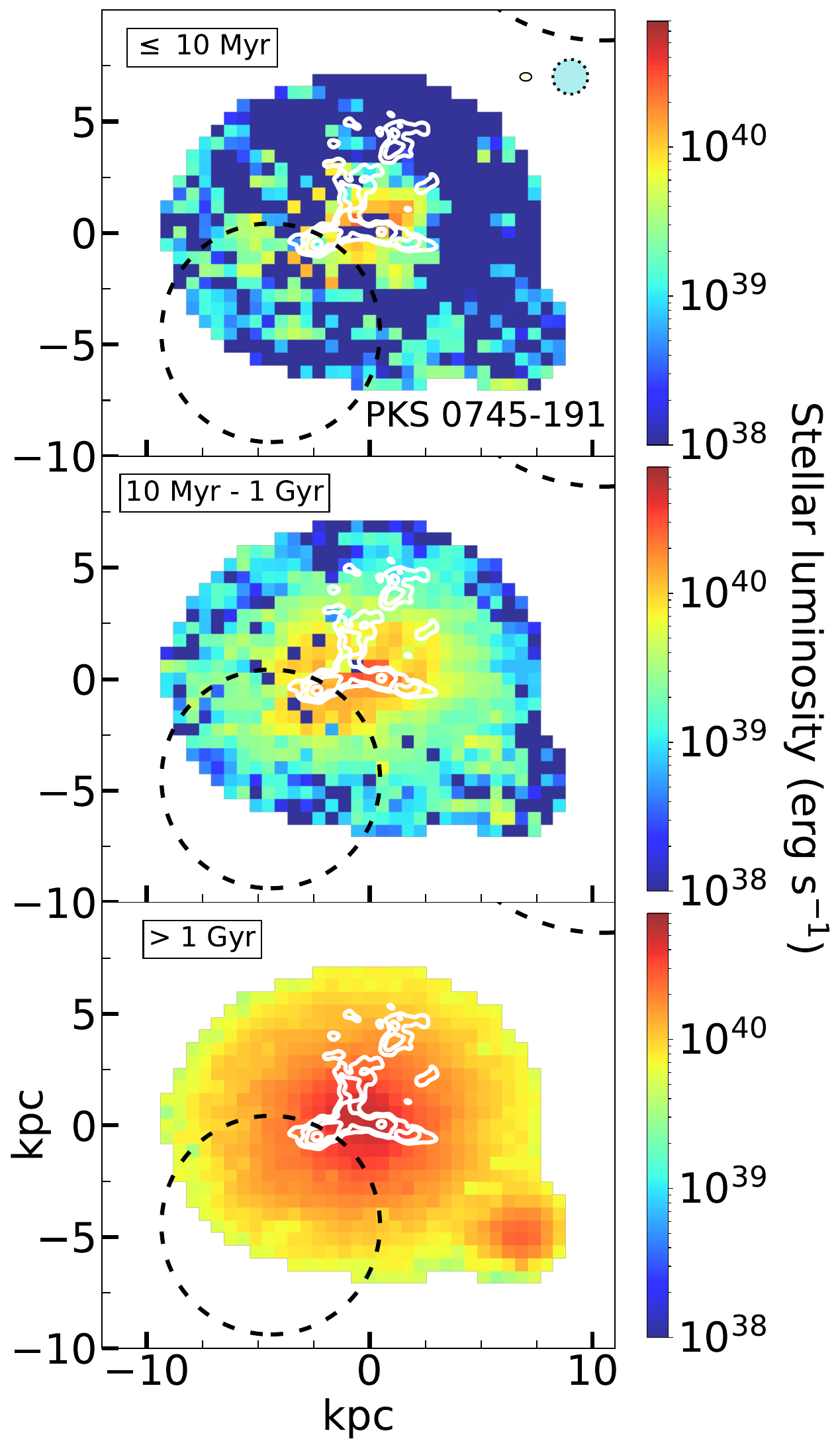}
    \caption{Stellar luminosity distributions in the central galaxy of PKS 0745$-$191 for young (top panel), intermediate-age (middle panel) and old (bottom panel) stars. The white contours show ALMA CO(3$-$2) emission flux, with contours of 10\%, 20\%, 40\%, and 80\% of the maximum CO(3$-$2) flux \citep{Russell2016}. In the top panel, the solid line (yellow interior) ellipse shows the beam size of the ALMA CO(3$-$2) observation and the dotted circle (cyan interior) shows the seeing of the KCWI observations. The dashed black ellipses show the positions of the X-ray cavities \citep{Sanders2014,Russell2019}.
    }
    \label{fig:PKS0745:Luminosities_ages}
\end{figure}

PKS 0745$-$191 is an archetypal cool cluster core with powerful radio emission \citep{Baum_ODea1991}, two known X-ray cavities, significant star formation rate of $\sim$ 17 M$_{\odot}$ yr$^{-1}$, and $\sim 5\times10^9$ M$_{\odot}$ of cold molecular gas distributed in three filaments \citep{ODea2008,Russell2016}.

\autoref{fig:PKS0745:flux;vstel;v50-vstel} maps the stellar flux, stellar velocity, and nebular gas velocity with respect to the stars. The stellar flux map shows two peaks corresponding to the central galaxy and a secondary galaxy. The galaxies are separated by a projected distance of $\sim 8.5$ kpc and with a radial velocity difference of 34 km s$^{-1}$. We detect stellar emission within a radius of $\sim 8$ kpc around the brightest spaxel with a total stellar luminosity of $1.8\times10^{43}$ erg s$^{-1}$.

We do not observe ordered motion in the $v_*$ map of PKS 0745$-$191, contrary to RX J0820.9+0752 and A1835. The range of velocities spanned by the stars is also smaller in the center of this galaxy than in the other two.
In addition, 
other than the bulk motion of the nebular gas with respect to the stars, no striking features in the $v_{50}^{\text{[OII]}}-v_*$ map are seen.

\autoref{fig:regions:PKS0745} shows the integrated spectra for the central and projected galaxies and the spaxels used for each spectrum. Their properties are listed in \autoref{tab:RXJ0820_regions}. The color excesses listed in \autoref{tab:RXJ0820_regions} indicate significant dust extinction in this system. Contrary to our other targets, PKS 0745$-$191 has substantial galactic extinction, with $E(B-V) = 0.45 \pm 0.01$ \citep{Galactic_Extinction1998}. As the $E(B-V)_*$ value for the secondary galaxy is less than what is expected from galactic extinction, the secondary galaxy has little to no dust. Nebular emission in the central galaxy is dusty with $E(B-V)_{\text{gas}} = 0.83\pm 0.08$ mag. This includes both the intrinsic and the galactic extinction. Even after removing the effects from galactic extinction, the intrinsic $E(B-V)_{\text{gas}}$ is $\sim 0.38$. The nebula in the central region of PKS 0745$-$191 is dusty, consistent with earlier observations \citep{Predehl1995}.

\autoref{fig:PKS0745:Ages_maps} maps the starlight fraction coming from young, intermediate-age, and old stars in the core of PKS 0745$-191$, while  \autoref{fig:PKS0745:Luminosities_ages} shows the luminosity emitted by young, intermediate-age, and old stars. These figures show that most of the starlight comes from stars older than 1 Gyr.  The top and central panels of \autoref{fig:PKS0745:Luminosities_ages} show higher luminosities from young and intermediate-age stars in the central region, roughly aligned along the southeast and northwest axis. This somewhat aligns with the X-ray cavities' axis and overlaps with the system's large molecular gas filaments \citep{Russell2016}, showed by the white contours in \autoref{fig:PKS0745:Luminosities_ages}.

The spectra in \autoref{fig:regions:PKS0745}, together with \autoref{fig:PKS0745:Ages_maps} and \autoref{fig:PKS0745:Luminosities_ages}, reveal significantly different stellar populations in the central galaxy (red) and the southwest projected galaxy (blue). The central galaxy shows ongoing star formation with nearly 40\% of its starlight coming from young or intermediate-age stars, whereas the projected galaxy is dominated almost entirely by old stars. As with A1835 and RX J0820.9+0752, the lack of young and intermediate-age stars in the projected galaxy suggests that the $\sim 5\times10^9$ M$_{\odot}$ reservoir of cold molecular gas in the core of PKS 0745$-$191 is unlikely to have been stripped from this galaxy.

Within our FOV the central galaxy has sustained an average star formation rate of $\sim 3$ M$_{\odot}$ yr$^{-1}$ over the past billion years, increasing to $\sim 8$ M$_{\odot}$ yr$^{-1}$ in the last 10 Myr. With a cold molecular gas reservoir of $\sim 5\times10^9$ M$_{\odot}$ \citep{Russell2016}, this implies a depletion timescale of roughly 625 Myr. As was the case for A1835, maintaining the current level of star formation in PKS 0745$-$191 requires that cooling replenishes the molecular gas reservoir.

\section{Summary}\label{sec:Discussion&Conclusion}

KCWI observations of the central galaxies in RX J0820.9+0752, A1835 and PKS 0745$–$191 offer new insights into AGN feedback and star formation in cool core clusters. By mapping the stellar continuum flux, line-of-sight stellar velocities, and stellar ages across the inner tens of kpc, we explore how starlight encodes the history of cooling and feedback processes in these three systems. 

These systems were selected for their strong cooling cores. Each shows distinct stellar dynamics and evolutionary histories. Our conclusions for each central galaxy can be summarized as follows:

\begin{itemize}
    \item RX J0820.9+0752: The system's blue arc is dominated by stellar emission from intermediate-age stars, challenging previous interpretations that photoionization from recent star formation caused the ionization of the cooling nebula. The old age of the stellar populations in both galaxies is inconsistent with the hypothesis that the system's massive reservoir originated from gas stripped from either galaxy. Instead, it is consistent with the gas cooling from the surrounding hot atmosphere, potentially triggered by gravitational effects of galaxy interactions.
    
    \item A1835: The inner region of the central galaxy of A1835 has significant star formation, with young and intermediate-age stars being detected up to $\sim 20$ kpc away from the nucleus. With almost $4\times 10^{10}$ M$_{\odot}$ of stars formed in the last Gyr, A1835 has long-lived star formation. East of the nucleus, we find a blueshifted clump of young stars which supports the presence of stimulated AGN feedback in this system. The kinematics of the stars and that of the nebular gas are consistent with stars cooling out of an outflow and raining back down towards the central AGN.

    \item PKS 0745-191: We focus on two galaxies in the center of PKS 0745-191. The secondary galaxy is almost exclusively comprised of stars older than 1 Gyr. The stellar population of the central galaxy is more nuanced, although most of the stellar emission also comes from old stars. Some young and intermediate-age stars are seen in the core, broadly overlapping with the known molecular gas filaments \citep{Russell2016}. Stellar emission from young stars is roughly aligned along the X-ray cavities' axis, indicating that AGN feedback likely plays a role in stimulating star formation in the core of PKS 0745-191.
\end{itemize}

All three systems contain significant reservoirs of cooling gas and show evidence of nearby galaxies within our field of view. However, these projected galaxies are dominated by old stellar populations, inconsistent with recent or ongoing star formation. This rules out the possibility that the observed cooling gas was stripped from these galaxies. Instead, our results support a scenario in which the cooling gas and subsequent star formation originated in situ, condensing directly from the hot intracluster medium.

In all three galaxies, there is spatial overlap between the youngest stars and the observed molecular gas, supporting the idea that star formation is fueled by this gas \citep{Pulido2018,Olivares2019,Russell2019}. 

RX J0820.9+0752 shows little to no recent star formation, with its blue arc dominated by intermediate-age stars (30–500 Myr). In contrast, A1835 and PKS 0745–191 have significant populations of young stars ($\leq$10 Myr) in their cores, suggesting more recent star formation activity. This difference may reflect varying feedback conditions: both A1835 and PKS 0745–191 host powerful AGN outbursts that can trigger localized cooling and star formation through cavity-driven uplift. The abundance of intermediate-age stars in RX J0820.9+0752 suggest an earlier event triggering cooling, possibly caused by galaxy interactions, with no evidence of ongoing or recent star formation. 

Their stellar kinematics also differ. A1835 shows coherent stellar motion with an east–west velocity gradient, while PKS 0745–191 lacks clear dynamical structure. This may reflect differences in stellar populations: A1835’s core is dominated by younger stars, which can retain kinematic signatures of recent feedback or interactions, while PKS 0745–191’s stellar emission is mostly from older stars that trace the bulk kinematics of the galaxy.

In A1835’s blueshifted clump, where young and intermediate-age stars dominate, we are effectively measuring the motion of recently formed stars. In PKS 0745$–$191, the overwhelming contribution from old stars masks any potential kinematic signature of younger populations. Future work using multi-component stellar velocity fitting methods will be needed to disentangle the kinematics of young stars in PKS 0745$–$191 and assess their role in AGN feedback-driven inflows or outflows.

\begin{acknowledgments}
The data presented here were obtained at the W. M. Keck Observatory, which is operated as a scientific partnership among the California Institute of Technology, the University of California, and the National Aeronautics and Space Administration. The observatory was made possible by the financial support of the W. M. Keck Foundation. We wish to recognize and acknowledge the very significant cultural role and reverence that the summit of Maunakea has always had within the indigenous Hawaiian community. We are most fortunate to have the opportunity to conduct observations from this mountain. We thank David Rupke for his invaluable help in analyzing the Keck observations using the IFSFIT pipeline. B.R.M acknowledges support from the Natural Sciences and Engineering Council of Canada and the Canadian Space Agency Space Science Enhancement Program. A.~L.~C. acknowledges support from the Ingrid and Joseph W. Hibben endowed chair at UC San Diego. The authors thank the anonymous referee for helpful comments that improved the paper.
\end{acknowledgments}

\facilities{Keck:II}

\software{Astropy \citep{Astropy1,Astropy2,Astropy3},  Python \citep{Python3}, Numpy \citep{Numpy1,Numpy2}, Matplotlib \citep{Matplotlib}, Scipy \citep{SciPy}, pPXF \citep{pPXF1,pPXF2}, IFSRED \citep{Rupke2014a}, IFSFIT \citep{Rupke2014b,IFSFIT2}
          }

\appendix

\section{Total FOV of A1835}
\label{appendix:A1835_galaxies}

\begin{figure}
    \centering
    \includegraphics[width=\columnwidth]{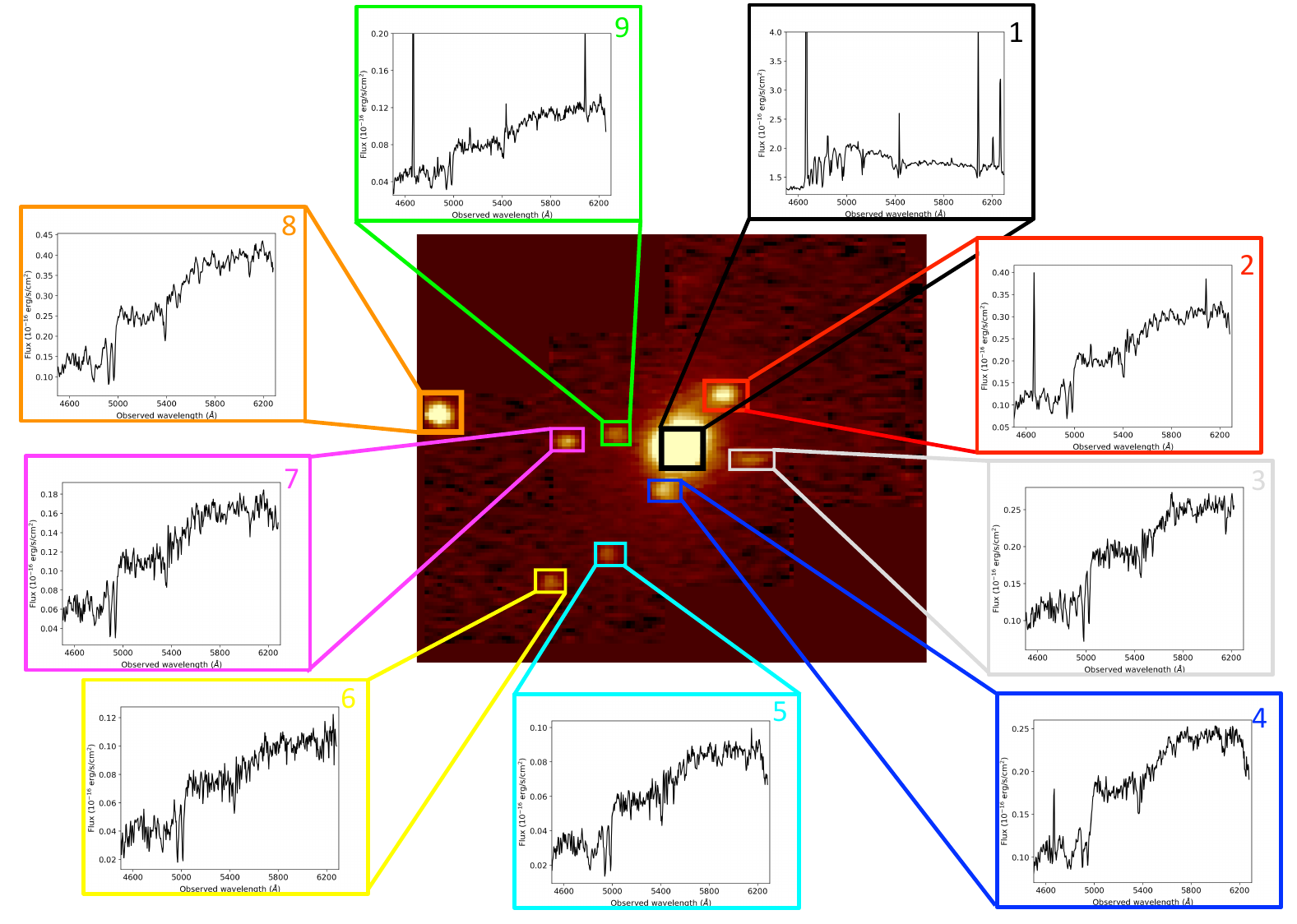}
    \caption{Total flux map of the mosaic image of A1835. Nine galaxies are present in the image, the BCG (1) and eight other galaxies which are projected along our line-of-sight (2$-$9). The integrated spectra for each boxed region are shown.}
    \label{fig:A1835_galaxies_spectrum}
\end{figure}

\autoref{fig:A1835_galaxies_spectrum} shows the total flux map for the total FOV observed for A1835. Apart from the central galaxy, we observed an additional eight projected galaxies. Each of the nine galaxies are identified in \autoref{fig:A1835_galaxies_spectrum} and the corresponding spatially-integrated spectrum is shown. The spaxels for each galaxy were summed and fitted using the technique described in \autoref{sec:Methods:Stellar_fit}. Most of the projected galaxies have old stellar populations with prominant Balmer breaks.

Since galaxy interactions can significantly affect galaxy dynamics, we compare the galaxies' positions and kinematics to investigate possible interactions with the central galaxy and its surrounding atmosphere. \autoref{tab:A1835_galaxies:table} lists the stellar redshifts measured from the resulting fits as well as listing each galaxy's relative velocity to the central galaxy. The percentage of the total starlight and the stellar luminosity within 1\arcsec{} around each galaxy's brightest spaxel are also listed. The galaxy numbers correspond to the numbers shown in the upper right of each integrated spectrum in \autoref{fig:A1835_galaxies_spectrum}.

The thousand km s$^{-1}$ velocity differences listed in \autoref{tab:A1835_galaxies:table} are inconsistent with interactions between the projected galaxies and the central galaxy. Only one projected galaxy, identified as Galaxy 8 in \autoref{fig:A1835_galaxies_spectrum}, has similar velocity as the central galaxy. While their stellar velocities differ by only $\sim 20$ km s$^{-1}$, their projected separation of more than 60 kpc makes tidal interactions between the two objects very unlikely. Moreover, since the projected galaxies have old stellar populations, it is unlikely that they are interacting with the central galaxy, as some star formation or younger stellar population would be expected in that case. Therefore, while our observations of the central galaxy of A1835 are accompanied by eight projected galaxies within our FOV, the central galaxy does not appear to be interacting with any of them. The A1835 cluster is known to have a velocity dispersion of $\sim 1500$ km s$^{-1}$ \citep{Czoske2004}, making most of the projected galaxies likely members of the cluster.

\begin{deluxetable}{ccccc}
        \tablecaption{Stellar velocities and luminosities of the galaxies in the FOV of our observations of A1835. \label{tab:A1835_galaxies:table}
        }
    \tabletypesize{\scriptsize}
    \tablewidth{0pt} 
    \tablehead{ \colhead{Galaxy}&\colhead{Redshift}&\colhead{Relative velocity}&\colhead{Stellar flux }&\colhead{Luminosity}\\ \colhead{} & \colhead{} & \colhead{(km/s)}&\colhead{(\%)}&\colhead{(erg/s)}}
\colnumbers
\startdata 
1&0.2513&0& 22.27&$4.63\times10^{43}$\\
2&0.2550&1113& 4.00&$8.34\times10^{42}$\\
3&0.2657&4304& 1.85&$3.85\times10^{42}$\\
4&0.2461&-1566& 3.15&$6.55\times10^{42}$\\
5&0.2557&1323& 0.91&$1.90\times10^{42}$\\
6&0.2625&3371& 0.99&$2.07\times10^{42}$\\
7&0.2440&-2189& 2.08&$4.32\times10^{42}$\\
8&0.2512&-20& 5.16&$1.07\times10^{43}$\\
9&0.2557&1321& 1.70&$3.53\times10^{42}$\\
\enddata
\tablecomments{
(1) Galaxy number, as indicated in \autoref{fig:A1835_galaxies_spectrum}.
(2) Redshift of the galaxy.
(3) Relative velocity of the galaxy with respect to the BCG.
(4) Percentage of the total starlight within a radius of 1\arcsec{} around the galaxy's brightest spaxel.
(5) Luminosity within a radius of 1\arcsec.}
\end{deluxetable}

\bibliography{sample631}{}
\bibliographystyle{aasjournal}

\end{document}